\documentclass[graybox]{svmult}

\usepackage{type1cm}        % activate if the above 3 fonts are
\usepackage{makeidx}         % allows index generation
\usepackage{graphicx}
\usepackage{multicol}        % used for the two-column index
\usepackage[bottom]{footmisc}% places footnotes at page bottom
\usepackage{newtxtext}       % 
\usepackage{newtxmath}       % selects Times Roman as basic font

\usepackage{braket}
\usepackage{amsmath}
\usepackage{mathrsfs}
\usepackage{mathtools}  
\usepackage{units}
\usepackage{color}
\usepackage{cite}

\begin{document}

%\preprint{}

\title{Attosecond Dynamics in Liquids}


\author{H. J. W\"orner, A. Schild, D. Jelovina, I. Jordan, C. Perry, T. T. Luu, Z. Yin}
\institute{H. J. W\"orner \at Laboratorium f\"ur Physikalische Chemie, ETH Z\"urich, 8093 Z\"urich, Switzerland\\ \email{hwoerner@ethz.ch} \and A. Schild, D. Jelovina, I. Jordan, T. T. Luu, Z. Yin \at Laboratorium f\"ur Physikalische Chemie, ETH Z\"urich, 8093 Z\"urich, Switzerland}

%\maketitle
%\date{\today}

%\begin{abstract}

%\end{abstract}

%\section*{abstract}

%\abstract{}

%\pacs{}
%\keywords{}

\maketitle

\vspace{-3.0cm}

\abstract{ 
Attosecond science is well developed for atoms and promising results have been obtained for molecules and solids. Here, we review the first steps in developing attosecond time-resolved measurements in liquids. These advances provide access to time-domain studies of electronic dynamics in the natural environment of chemical reactions and biological processes. We concentrate on two techniques that are representative of the two main branches of attosecond science: pump-probe measurements using attosecond pulses and high-harmonic spectroscopy (HHS). In the first part, we discuss attosecond photoelectron spectroscopy with cylindrical microjets and its application to measure time delays between liquid and gaseous water. We present the experimental techniques, the new data-analysis methods and the experimental results. We describe in detail the conceptual and theoretical framework required to fully describe attosecond chronoscopy in liquids at a quantum-mechanical level. This includes photoionization delays, scattering delays, as well as a coherent description of electron transport and (laser-assisted) photoemission and scattering. As a consequence, we show that attosecond chronoscopy of liquids is, in general, sensitive to both types of delays, as well as the electron mean-free paths. Through detailed modeling, involving state-of-the-art quantum scattering and Monte-Carlo trajectory methods, we show that the photoionization delays dominate in attosecond chronoscopy of liquid water at photon energies of 20-30~eV. This conclusion is supported by a near-quantitative agreement between experiment and theory. 
In the second part, we introduce HHS of liquids based on flat microjets. These results represent the first observation of high-harmonic generation (HHG) in liquids extending well beyond the visible into the extreme-ultraviolet regime. We show that the cut-off energy scales almost linearly with the peak electric field of the driver and that the yield of all harmonics scales non-perturbatively. We also discuss the ellipticity dependence of the liquid-phase harmonics, which is systematically broadened compared to the gas phase. We introduce a strongly-driven few-band model as a zero-order approximation of HHG in liquids and demonstrate the sensitivity of HHG to the electronic structure of liquids. Finally, we discuss future possibilities for modelling liquid-phase HHG, building on the methods introduced in the first part of this chapter. 
In the conclusion, we present an outlook on future studies of attosecond dynamics in liquids.
}

%\tableofcontents

\section{Introduction}

Attosecond science has the potential to address fundamental open questions in chemical and physical sciences. By directly accessing the electronic dynamics of matter on a time scale that naturally freezes any structural dynamics, attosecond spectroscopy targets the most fundamental electronic processes that define the properties of matter. The methods of attosecond science have been well established and most broadly demonstrated on atoms \cite{drescher02a,goulielmakis10a,kluender11a,ott13a}. Very promising results have also been obtained on molecules \cite{niikura02a,baker06a,calegari14a,kraus15b,huppert16a} and solids \cite{cavalieri07a,neppl15a,tao16a,luu15a,vampa15a}. The field of attosecond molecular dynamics has recently been reviewed in Ref. \cite{baykusheva20a}.

Attosecond dynamics in liquids have so far not been accessible to time-resolved experiments. However, they are expected to play a particularly important role in chemical and biological processes that dominantly take place in the liquid phase. The primary processes underlying charge and energy transfer in the liquid phase take place on few-femtosecond to attosecond time scales \cite{may11a} and have eluded direct experimental access to date. These processes include e.g. the ionization of liquid water, electron scattering during its transport through the liquid phase, as well as intermolecular Coulombic decay (ICD) \cite{cederbaum97a} and electron-transfer-mediated decay (ETMD) \cite{zobeley01a}, to name only a few examples.

Among these, the ionization of liquid water plays a particularly important role since it initiates the dominant pathway of radiation damage in living matter (see Ref. \cite{nikjoo16a} for a recent review). The ionization of liquid water is inherently an attosecond process and therefore offers a target of primary relevance for attosecond science. However, the primary ionization event itself is not the only attosecond process involved in the ionization of liquid water. The subsequent electron scattering during its transport through the liquid phase also takes place on a sub-femtosecond time scale. It controls the energy deposition during electron scattering and thereby the chemistry initiated by ionizing radiation in liquid water. Important new insights into the early femtosecond time scale of water ionization have very recently been obtained using both one-photon ionization \cite{svoboda20a} and strong-field ionization \cite{loh20a}.

In this book chapter, we discuss two recently demonstrated experimental approaches to attosecond science in the liquid phase: attosecond photoelectron spectroscopy \cite{jordan20a} and high-harmonic spectroscopy \cite{luu18a}. We present the experimental methodology and discuss the first experimental results, along with the developed conceptual framework and the related theoretical methods. 

In the first part, we describe the principles and methods of attosecond photoelectron spectroscopy of liquids \cite{jordan20a}. This technique has been applied to measure the time delay between photoemission from liquid and gaseous water. Photoionization delays from isolated molecules in the gas phase have only recently been measured \cite{huppert16a}. These measurements have been interpreted through the development of a general theoretical and computational methodology \cite{baykusheva17a}. Here, we use photoemission from gaseous water molecules as a reference against which photoemission from liquid water is clocked. We find that photoemission from liquid water is {\it delayed} by $\sim$50-70 attoseconds compared to photoemission from isolated water molecules at photon energies of $\sim$20-30 eV. We show through detailed modelling that includes quantum-scattering calculations on water clusters of variable size, as well as Monte-Carlo trajectory simulations benchmarked against the time-dependent Schr\"odinger equation, that the measured delays are dominated by the local solvation structure in liquid water. Electron scattering during transport, in contrast, makes a negligible contribution to the measured delays.

In the second part, we describe the experimental methods and first results on non-perturbative high-harmonic generation in liquids \cite{luu18a}. This recent discovery represented a paradigm change in attosecond science because a series of previous experimental works on water droplets \cite{flettner03a,kurz13a,kurz16a} concluded that high-harmonic generation was impossible at the natural density of liquid water. Previous observations of harmonic generation from liquids were indeed limited to low-order harmonics in the visible range \cite{dichiara09a} and to the coherent-wake-emission regime \cite{heissler14a}, where high-harmonic generation takes place in a plasma created by the ultra-intense laser pulse (peak intensity $> 10^{18}$ W/cm$^2$) rather than in the original target. The main observations from our work \cite{luu18a} are a linear scaling of the cut-off energy with the electric field, a pronounced sensitivity of the observed spectra on the electronic structure of the liquids and a broadened ellipticity dependence of the high-harmonic yield compared to gas-phase HHG. High-harmonic spectroscopy of liquids offers a complementary and remarkably different approach to the same fundamental problems discussed above. It provides access to the attosecond dynamics of the ionization step. It is sensitive to the dynamics of electron transport through the liquid phase, and thereby to electron scattering dynamics. Finally, it might also be sensitive to the spatial shape of the created electron hole through the recombination step. Liquid-phase high-harmonic spectroscopy therefore offers several promising prospects for exploring the electronic structure and dynamics of liquids, solutes and solvents.

In the outlook that concludes this book chapter, we discuss possible future avenues of attosecond science in liquids. These include time-resolved studies of ICD and ETMD, which are the prototype elementary processes of charge and energy transfer in living matter, but have unknown time scales. 

\section{Attosecond photoelectron spectroscopy of liquid water}

The most direct way to probe electron dynamics in liquid water is to perform a time-resolved measurement on the attosecond time scale. For this purpose, we have chosen attosecond photoelectron spectroscopy. It consists in creating freely moving electrons in the conduction band of liquid water through ionization with an attosecond pulse and probing their dynamics through interaction with a temporally delayed near-infrared (NIR) pulse before detecting them through photoelectron spectroscopy. 

\begin{figure}[h!]
\begin{center}
\includegraphics[width=\textwidth]{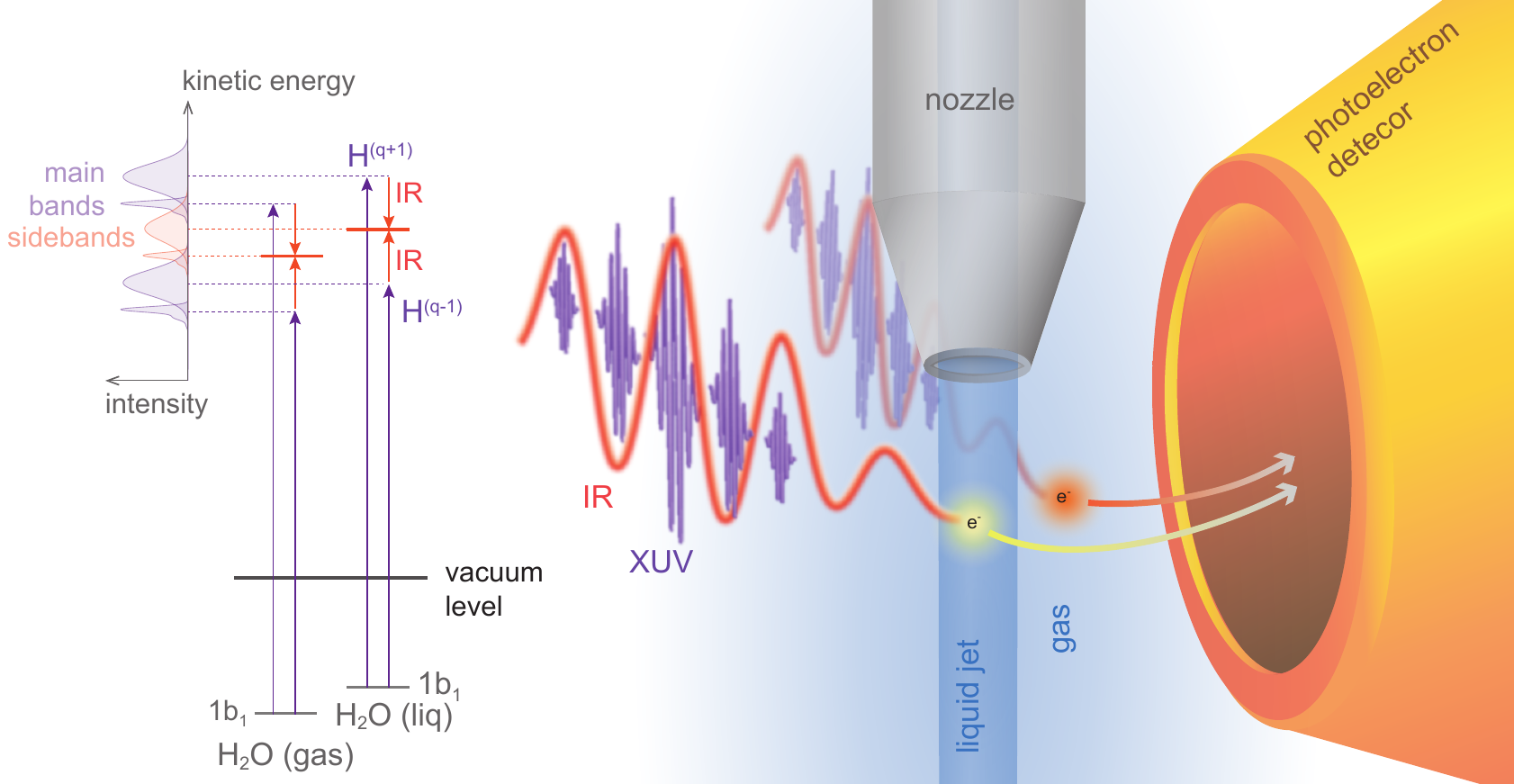}
\end{center}
\caption{{\bf Attosecond time-resolved photoelectron spectroscopy of liquid water}. A spectrally filtered attosecond pulse train, composed of a few high-harmonic orders (H$^{\rm (q-1)}$,H$^{\rm (q+1)}$, etc.) superimposed with a near-infrared femtosecond laser pulse interacts with a microjet of liquid water. Photoelectrons are simultaneously emitted from the liquid and the surrounding gas phase. The resulting photoelectron spectra are measured as a function of the time delay between the overlapping pulses. Adapted from Ref. \cite{jordan20a}.}
\label{concept}
\end{figure}

For the first experiment, described herein, we have used attosecond interferometry, which combines an extreme ultraviolet (XUV) attosecond pulse train (APT) with a femtosecond NIR laser pulse (see Fig. \ref{concept}). This method has previously been applied to measure photoionization delays in atoms \cite{kluender11a} and molecules \cite{huppert16a}. The particular advantage of attosecond interferometry over e.g. attosecond streaking is the high spectral resolution that is achieved \cite{isinger17a}, in addition to the high temporal resolution that is common to both methods. Whereas the spectral resolution in attosecond streaking is limited by the temporal duration of the attosecond pulse through the time-bandwidth product, the spectral resolution in attosecond interferometry is not. Instead, the resolution is limited by the spectral width of the individual harmonic orders, which are limited by the length of the APT and its chirp over the (femtosecond) time scale of the APT. The high spectral resolution is a crucial aspect of the present experiments because it enables the spectral distinction of photoelectrons originating from the gas and liquid phases. These give rise, respectively, to the narrow and broad spectral features illustrated in Fig. \ref{concept}.

Experiments based on attosecond interferometry or attosecond streaking directly measure time delays in photoionization, but all such measurements are inherently relative at present. It is therefore essential to identify a suitable reference against which these delays can be measured. In the present experiment, we turn the presence of the evaporating gas-phase molecules, usually perceived as a nuisance in liquid-microjet photoelectron spectroscopy, into an advantage. By measuring photoionization time delays of liquid water relative to those of isolated water molecules, we eliminate the ''trivial'' single-molecule contributions to the photoionization delays. The word ''trivial'' is consciously put in quotation marks because the measurement and interpretation of molecular photoionization delays is itself a highly active and intriguing area of research \cite{huppert16a,baykusheva17a,vos18a,cattaneo18a,biswas20a}. However, the goal of the present work goes much beyond this, by aiming at identifying the specific contributions of the liquid phase to the attosecond photoemission delays. These include two types of effects, i.e. the modification of the electronic structure and therefore the scattering potential of an isolated water molecule caused by solvation, as well as electron scattering dynamics during its transport through the liquid phase. 

In the following sections, we discuss the experimental techniques, which include the realization of liquid-phase attosecond interferometry and the methods developed to retrieve photoionization delays from overlapping photoelectron spectra. We also discuss in detail the novel theoretical methods, which comprise the integration of laser-assisted scattering in the formalism of attosecond interferometry, the development of a general three-dimensional Monte-Carlo code for simulating such experiments and the identification of condensation effects on the photoionization delays from water clusters. This extensive development of experimental and theoretical methods has culminated in the quantitative interpretation of the measured photoemission time delays of liquid water.

\subsection{Experimental methods and results}
\label{exp}

\begin{figure}[h!]
\begin{center}
\includegraphics[width=\textwidth]{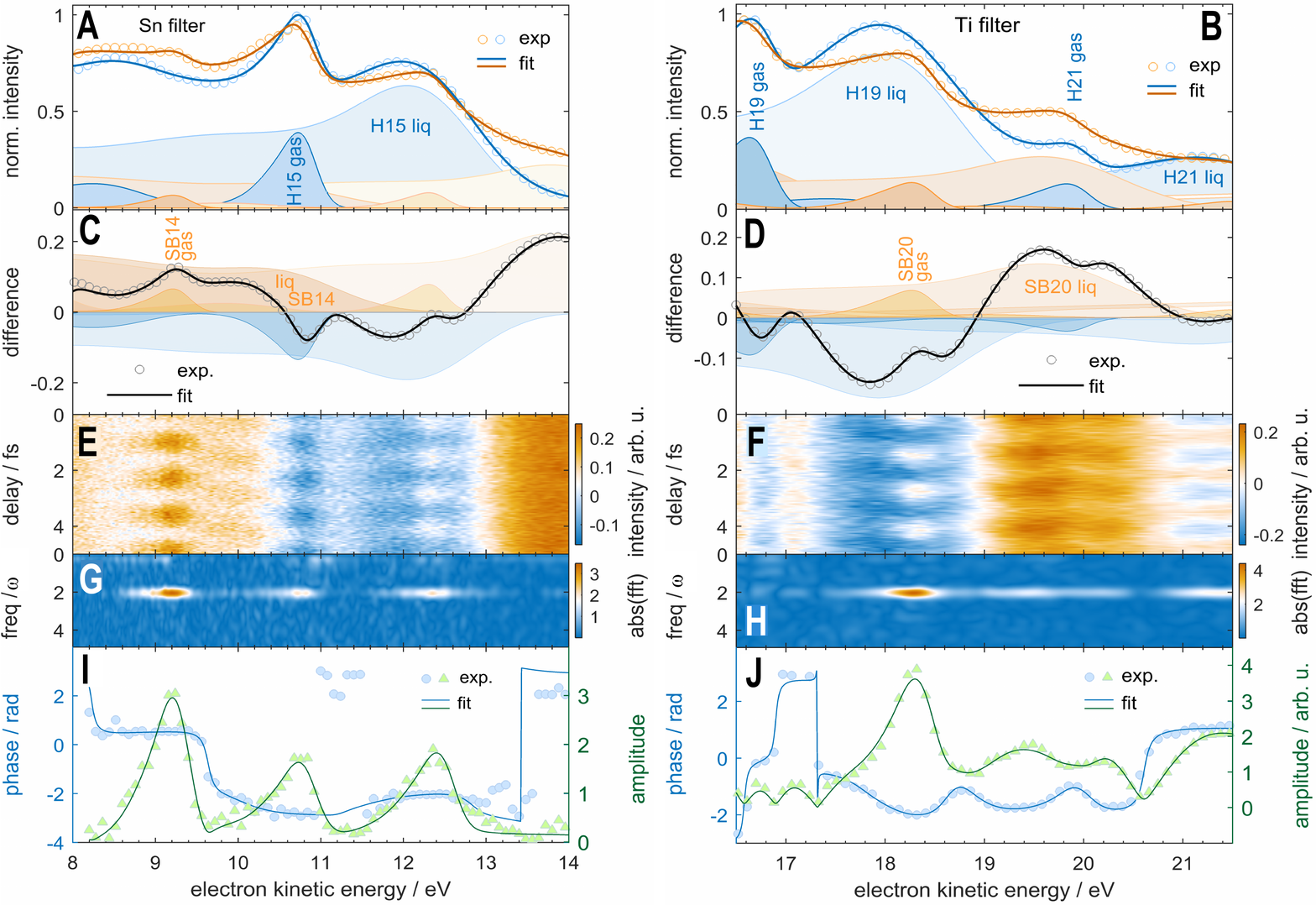}
\end{center}
\caption{{\bf Attosecond photoelectron spectra of liquid and gaseous water}. Data acquired with a Sn-filtered APT (left) or a Ti-filtered APT (right). a) Photoelectron spectra in the absence (blue) and presence (orange) of the IR field with their principal-component fit (full lines) and decomposition (filled curves). b) Difference spectra (circles), principle-component fit (line) and decomposition (filled curves) into side bands (orange) and depletion (blue). c) Difference spectra as a function of the APT-IR time delay. d) Fourier-transform power spectrum of c. e) Amplitude and phase of the 2$\omega$ component of the Fourier transform. Adapted from Ref. \cite{jordan20a}.}
\label{data}
\end{figure}

The experimental setup consists of an attosecond beamline delivering XUV APTs and a femtosecond NIR pulse, a liquid microjet delivering a micrometer-thin stream of liquid into a high-vacuum chamber, and a photoelectron spectrometer. In this work, we have used the attosecond beamline described in \cite{huppert15a}, which uses nested white-light and monochromatic interferometers to actively stabilize the time delay between NIR and XUV pulses to extreme accuracy. High-harmonic generation from a NIR laser pulse centered at 800 nm in an argon gas cell is used to generate the APT. The created APT is spectrally filtered by thin metallic foils to reduce its spectral bandwidth and thereby the spectral overlap in the photoelectron spectra. Specifically, we have used Sn filters to isolate the harmonic orders 11, 13 and 15, or Ti filters to isolate the harmonic orders 17, 19 and 21. A particular advantage of these metal filters is their sharp spectral truncation at the high-energy side of the spectrum, which makes the remaining spectral overlap manageable (see Fig. \ref{data}). The liquid microjet is formed by expanding high-purity liquid water (with the addition of NaCl to a concentration of 50 mM) through a $\sim$20~$\mu$m inner-diameter capillary with a high-performance liquid-chromatography pump. The photoelectron spectrometer is a 1-m long magnetic-bottle time-of-flight spectrometer previously described in Ref. \cite{jordan15a}.

The experimental results are shown in Fig. \ref{data}, where the left (right) columns show the data acquired with the Sn (Ti) filter. Correspondingly, the data on the left-hand side is mainly composed of photoelectron spectra generated by H13 and H15, whereas the data on the right-hand side mainly originates from H19 and H21. These data have been analyzed through a principal-component analysis (PCA) based on photoelectron spectra of gaseous and liquid water that were measured with a monochromatized high-harmonic light source \cite{vonconta16a} and the same photoelectron spectrometer \cite{jordan15a}. This PCA provides a unique decomposition of the measured spectra into its constituents. The panels a) in Fig. \ref{data} show photoelectron spectra acquired with the XUV APT only (blue) or with XUV-APT and NIR pulses (orange). The panels b) show the difference spectra, obtained by subtracting the XUV-only spectra from the XUV+NIR spectra. In the panels a) and b) the circles represent the experimental data and the full lines the PCA fit. The contributing principal components are shown as filled areas, whereby the XUV-only contributions are shown in blue and the side-band contributions are shown in orange. The perfect agreement between the PCA fit and the experimental data shows that the measured spectra have been successfully decomposed into the contributing photoelectron spectra from liquid and gaseous water of the relevant high-harmonic orders.

The panels c) show the difference spectra (from panels b) as a function of the time delay between the overlapping XUV-APT and NIR pulses. Positive signals (in orange) are dominated by side-band contributions and negative signals (in blue) are dominated by the depletion of the XUV-APT photoelectron spectra. The Fourier transform of the data in panels c) along the time-delay axis is shown in panels d) and the phase and amplitude of the Fourier transform, integrated over the dominant 2$\omega$ frequency component is shown in panels e). These complex-valued Fourier transforms have been analyzed within the PCA with minimal assumptions, i.e. the entire photoelectron bands corresponding to the highest-occupied molecular orbital (HOMO) of the gas or liquid contributions are assigned a characteristic phase ($\phi_{\rm gas}$ or $\phi_{\rm liq}$). This minimal assumption is sufficient to reproduce the complete electron-kinetic-energy-dependent complex-valued Fourier transforms, as illustrated in panels e. As a consequence, it is possible to determine the photoemission time delays between the HOMO bands of the liquid and gas phases:
\begin{equation}
\Delta\tau=\tau_{\rm liq}-\tau_{\rm gas}=\frac{\phi_{\rm liq}-\phi_{\rm gas}}{2\omega}.
\end{equation}
We find $\tau_{\rm liq}-\tau_{\rm gas}=69\pm 20$~as at a photon energy of 21.7 eV (SB14) and $\tau_{\rm liq}-\tau_{\rm gas}=48\pm 16$~as at a photon energy of 31.0 eV (SB20).

In addition to robust delays, the PCA also enables the determination of reliable modulation depths of the side-band intensities. This additional observable of attosecond interferometry has received very little attention to date. In an idealized situation where the amplitude of the two-photon-ionization pathways leading to the same side-band state have the same amplitude, the modulation contrast will be 100 \%. The experimentally observed modulation contrast will in general be smaller than 100 \% for various possible reasons. In the gas phase, these include i) different amplitudes of the two-photon-ionization pathways, ii) contributions of several two-photon-ionization pathways that do not interfere, e.g. because they correspond to different initial eigenstates of they leave the system in different final states, iii) contribution of ionization channels (e.g. emission angle or target orientation) corresponding to different time delays, iv) fluctuations in XUV-IR path-length difference, etc. In condensed-phase attosecond interferometry, additional effects include v) contributions from different local environments around the photoionized entity and vi) decoherence of the photoelectron wave packet during transport through the condensed phase.

Given the many possible contributions to a finite modulation contrast, it is thus important to find a meaningful reference. In the present case, the best reference is our previously published measurement of molecular photoionization delays in water vapor \cite{huppert16a}, which was carried out with the same apparatus, under the same experimental conditions as the liquid-phase experiments. This experiment revealed a modulation contrast of HOMO SB14 indistinguishable from 100 \% within the experimental signal-to-noise ratio (see Fig. 4.3 in \cite{huppert16b}). 

Using the PCA discussed above, we have determined a relative modulation depth of the liquid- compared to the gas-phase signal $M_{\rm r}=M_{\rm liq}/M_{\rm gas}$ of $0.17\pm 0.03$ at a photon energy of 21.7~eV and $0.45\pm 0.06$ at a photon energy of 31.0~eV. The determination of relative modulation depths has the advantage that it eliminates the contributions i)-iv), at least as far as single-molecule contributions and the experimental imperfections are concerned. Since a deviation from a perfect modulation contrast was not observable in water vapor \cite{huppert16b}, the finite modulation depths reported above can mainly be attributed to the effects v) and vi). As further discussed in Section \ref{interpr}, contribution v) is significant and contribution vi) might also be important.

\subsection{Concepts and theoretical methods}
\label{concepts}

In this section, a detailed theoretical treatment of attosecond photoelectron spectroscopy (APES) in liquids is given. Although a significant amount of work on APES of solids has previously been published (see e.g. \cite{zhang09a,liao14b,borisov13a,pazourek15a}), a comprehensive treatment that includes both photoionization delays and scattering delays has been missing until our recent work \cite{rattenbacher18a}. Motivated by experimental results \cite{cavalieri07a,neppl12a,neppl15a,okell15a,locher15a,tao16a,siek17a}, most previous theoretical treatments were designed to describe experiments on metallic samples. The description of APES in liquids fundamentally differs from the situation encountered in metals. This is due to the different penetration depths of the XUV and NIR fields. Whereas the XUV fields penetrate more deeply into metals than the NIR, the situation is reversed in liquids, which are practically transparent to the NIR radiation. Since the temporal information originates from the interaction of the XUV-induced photoelectron wave packet with the NIR radiation, APES on metals is mainly sensitive to the transport time of the photoelectron wave packet from the point of ionization to the metal-vacuum interface. In APES of liquids, the situation is reversed and the NIR field is present throughout the medium. Therefore APES in liquid is sensitive to other aspects of attosecond photoionization dynamics. The following sections present a detailed conceptual and theoretical analysis of this situation.

\subsubsection{Time delays in photoionization and scattering}

Time delays in quantum scattering were first analyzed by Wigner and Smith \cite{wigner55a,smith60a}. They studied conventional scattering events, corresponding to the collision of two particles. Photoionization can be understood as a half collision and can be described  with a very similar quantum-mechanical formalism. Consequently, it is not surprising that photoionization delays are closely related to scattering delays, both conceptually and in terms of their magnitudes \cite{dahlstrom12a,pazourek15a}. 

Both scattering and photoionization delays typically lie in the attosecond domain. They have therefore become accessible in the time domain only recently \cite{schultze10a,kluender11a}. Attosecond science has so far focused on the measurement of time delays in photoionization because they are directly accessible to APES. As a consequence, time delays in (conventional) scattering, i.e. full collisions have so far not been discussed. Here, we show that they can play a significant role in condensed-phase APES. More importantly, we show that laser-assisted electron scattering is a general phenomenon that needs to be included into a comprehensive treatment of condensed-phase  APES.

\begin{figure}[h!]
\begin{center}
\includegraphics[width=0.8\textwidth]{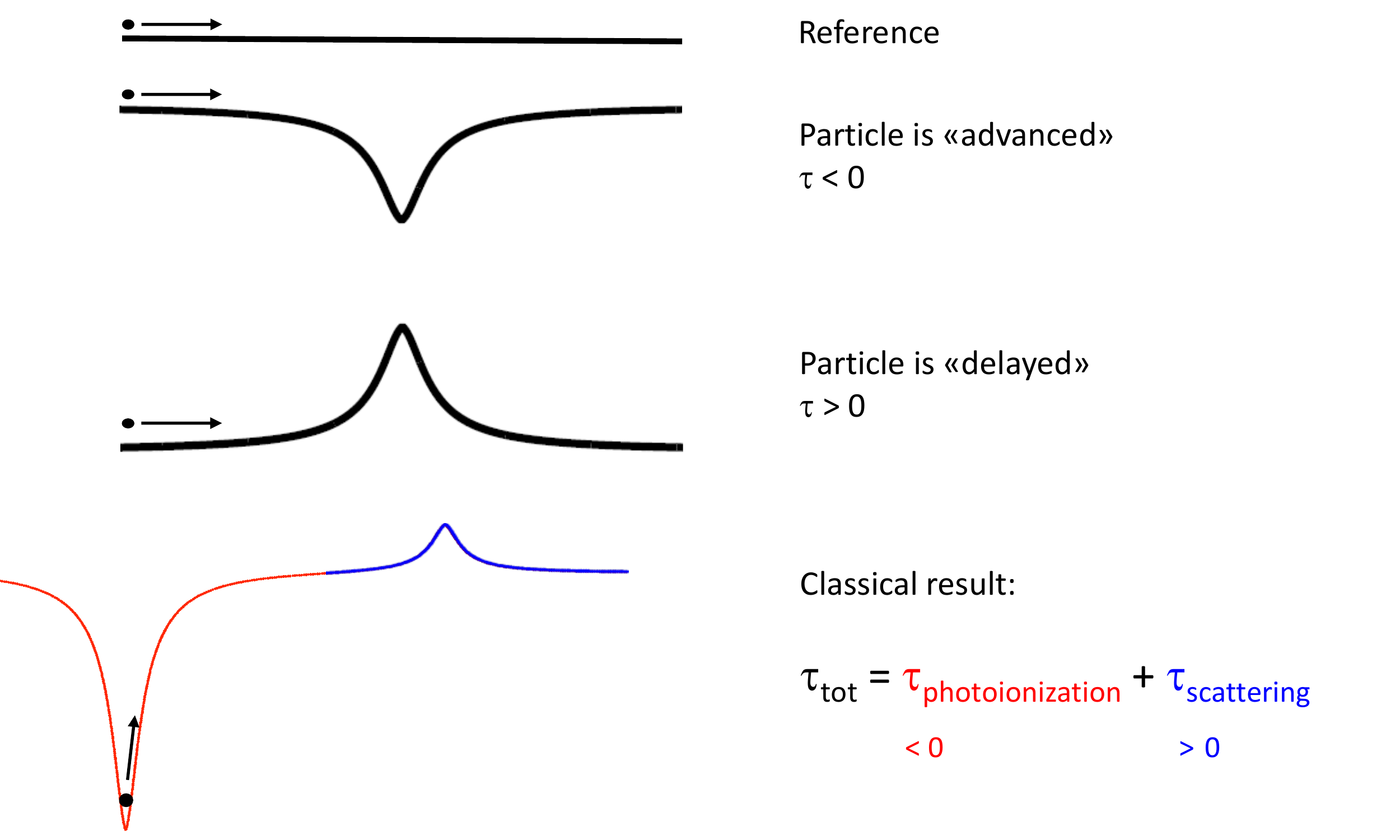}
\end{center}
\caption{{\bf Time delays in photoionization and scattering in the classical limit}. The time delays are defined as the differences in the arrival times of the classical particles that have propagated through different potentials with that of a ''reference'' particle propagating through potential-free space with the same final (asymptotic) velocity.}
\label{classical}
\end{figure}

We initiate our discussion with a classical analysis that provides a transparent basic understanding of the problem. Figure \ref{classical} illustrates the definition of these classical delays. It is easy to understand that an attractive (repulsive) potential results in a negative (positive) delay because the particle is locally accelerated (decelerated) compared to the potential-free propagation before returning to the same asymptotic velocity. As a consequence the time required to cover the same horizontal distance is smaller (larger) compared to potential-free propagation, which results in a negative (positive) delay, as illustrated. In these classical considerations, it is also easy to understand that such classical scattering delays are additive. As long as the potentials do not overlap, the delay contributed by a particular potential is only a property of the local potential and does not depend on the previous trajectory of the particle. 

The same considerations hold true for a classical treatment of photoionization delays. This situation is illustrated at the bottom of Fig. \ref{classical}. The classical definition of a photoionization delay corresponds to a situation where the particle starts from the bottom of an attractive potential with a finite velocity. This velocity must be chosen such that the final (asymptotic) velocity of the particle is the same as for the reference particle. This classical photoionization delay will be negative for the same reason as the classical scattering delay for an attractive potential is negative. The classical photoionization delay is equal to half of the classical scattering delay for the same potential. 

Continuing on this analogy, it is easy to show that the total classical delay for one photoionization and one scattering event is simply the sum of the two classical delays, as long as the potentials do not overlap. This situation is illustrated at the bottom of Fig. \ref{classical}.

\subsubsection{Non-local attosecond interferometry in one dimension}
\label{NL1D}

The translation of the results from the previous section to observables of attosecond spectroscopy requires at least two modifications. First, electrons are not classical objects and should therefore be described quantum-mechanically. When this is not possible, the classical approximation has to be justified, as we further discuss below. Second, the situation analyzed in classical terms in the previous section would correspond to a time-of-flight measurement with attosecond resolution, which is impractical. Instead, attosecond spectroscopy measures such time delays through two-color photoionization. Whereas the relation of photoionization delays measured with such techniques with the (quantum-mechanical) Wigner delays has been established \cite{dahlstrom12a,pazourek15a}, the treatment of full-scattering events itself and full-scattering in combination with photoionization has only been addressed in our recent publication \cite{rattenbacher18a}.

In this section, we present a complete quantum-mechanical analysis of attosecond time delays caused by photoionization and scattering as measured by attosecond interferometry. We note that these results are not specific to attosecond interferometry, but also apply to attosecond streaking, which will be the subject of future publications. Starting from the numerical solution of the time-dependent Schr\"odinger equation, we show that the obtained delays remarkably deviate from the classical expectation. Building on analytical descriptions of laser-assisted photoemission (LAPE) and laser-assisted electron scattering (LAES), we develop a complete analytical description of the quantum-mechanical problem at hand. This has led to the discovery of a novel phenomenon that we call ''non-local'' attosecond interferometry. In contrast to the traditional understanding of attosecond interferometry, where the interaction of the photoelectron wave packet with the XUV and NIR radiation takes place in a common spatial volume, the non-local pathways involve interactions at spatially separated regions. Specifically, the XUV absorption is localized to the position of the initially bound electronic wave function, whereas the NIR interaction takes place at the position where scattering occurs. These ''non-local'' XUV/NIR interactions lead to the interesting result that the total delays measured in the presence of photoionization and scattering are, in general, sensitive to both photoionization and scattering delays, and to the distances travelled between XUV and NIR interactions, i.e. to the mean-free paths of electron scattering.

\begin{figure}[h!]
\begin{center}
\includegraphics[width=0.8\textwidth]{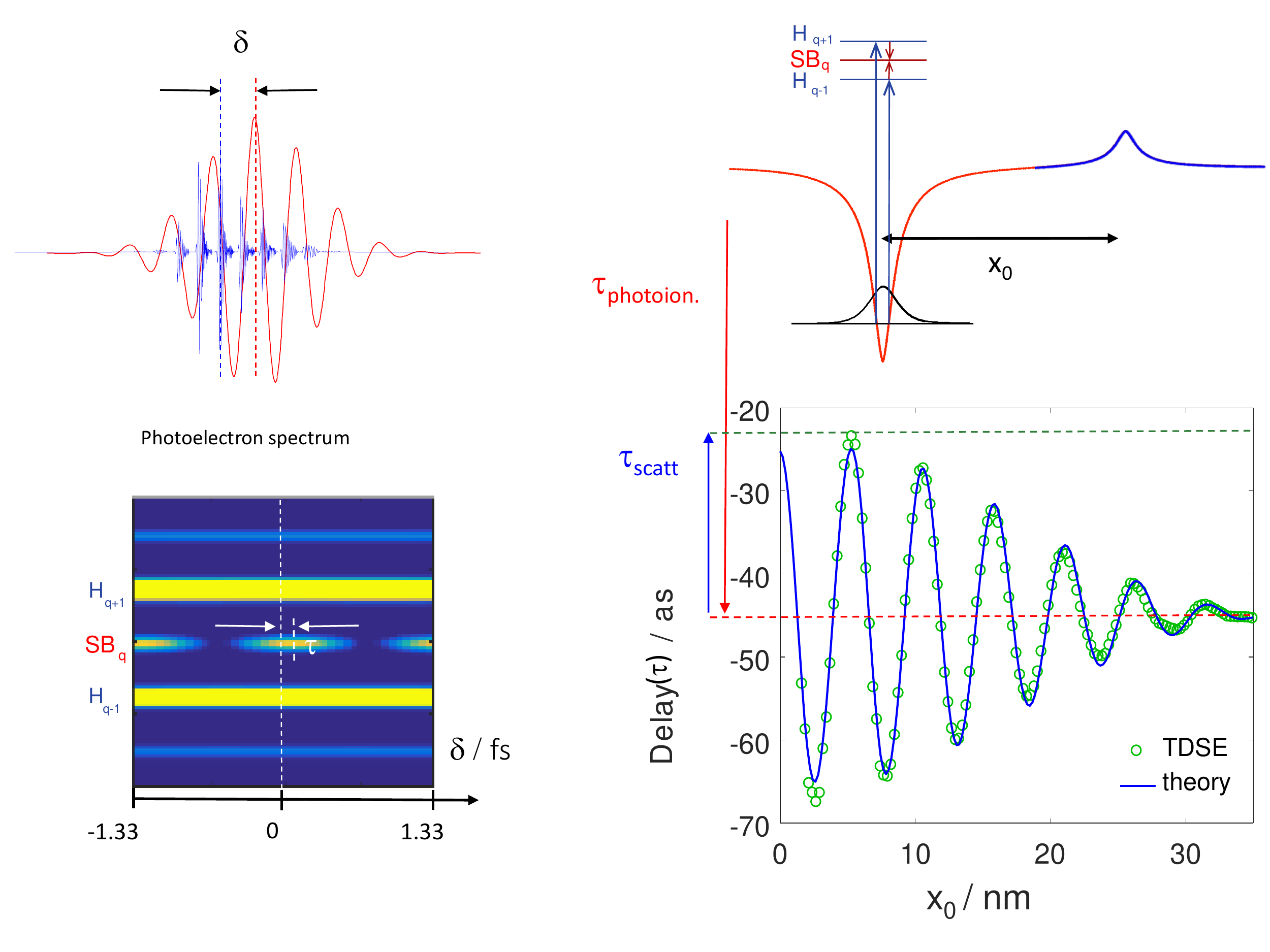}
\end{center}
\caption{{\bf Time delays in photoionization and scattering as measured by attosecond interferometry}. This figure shows (schematically) the employed XUV and NIR fields (top left), the used model potential (top right), the calculated photoelectron spectrum as a function of the time delay $\delta$ and the calculated delay $\tau$ as a function of the separation $x_0$ between the attractive and repulsive potentials.}
\label{model}
\end{figure}

We solve the one-electron time-dependent Schr\"odinger equation (TDSE) in one dimension using a model potential consisting of an attractive and a repulsive potential well (see Fig. \ref{model}). Details of this calculation are given in Ref. \cite{rattenbacher18a}. The system is initiated in the lowest bound electronic state of the attractive potential. The TDSE is solved in the presence of an XUV field consisting of two harmonic orders (H$_{q-1}$ and H$_{q+1}$) and a NIR field. The photoelectron spectrum is calculated by projecting the total wave function at the end of the propagation onto the continuum eigenstates of the model system. This gives rise to a photoelectron spectrum as shown on the bottom left of Fig. \ref{model}. The calculated photoelectron spectra reveal an oscillation of the intensity of SB$_q$ as a function of the delay $\delta$ between the XUV and NIR pulses. In the calculation we actually vary the carrier-envelope phase of the NIR pulse instead of varying the time delay $\delta$. This is equivalent for the present purpose but avoids unwanted effects caused by the envelope functions of the two fields.  The temporal shift of the SB$_q$ maximum with respect to $\delta =0$ defines the measured time delay $\tau$.

Solving the TDSE under the described conditions for different separations $x_0$ between the attractive and repulsive potentials provides the result shown in the bottom-right part of Fig. \ref{model}. The total delay displays an oscillatory dependence on $x_0$. 
This is in remarkable contrast to the naive classical expectation discussed in the previous section. For our model system, the photoionization delays amounts to -44~as, as indicated by the red dashed line, and the scattering delay to +21~as. Within our classical considerations this would lead to a total delay of -23~as, independent of $x_0$ as indicated by the dashed green line. The result of the TDSE (green circles) departs markedly from this classical prediction. The calculated delay $\tau$ is found to oscillate around the value of the photoionization delay with a maximal amplitude given by the scattering delay. More precisely, the total delay $\tau$ can be expressed as a sum of the photoionization delay $\tau_{\rm PI}$  and a non-local delay $\tau_{\rm nl}$,
    \begin{align}
      \tau = \tau_{\rm PI} + \tau_{\rm nl},
      \label{eq:tausum}
    \end{align}
    where $\tau_{\rm nl}$ oscillates with the distance $x_0$.

This result significantly contrasts with the previous understanding of attosecond interferometry on isolated particles, where only local XUV and NIR interactions are important. In such a framework, the observed oscillations cannot be explained. They could, however, be quantitatively explained in Ref. \cite{rattenbacher18a} by developing an analytical formalism that includes both local and non-local pathways. 

    \begin{figure}[htbp]
      \centering
      \includegraphics[width=0.8\textwidth]{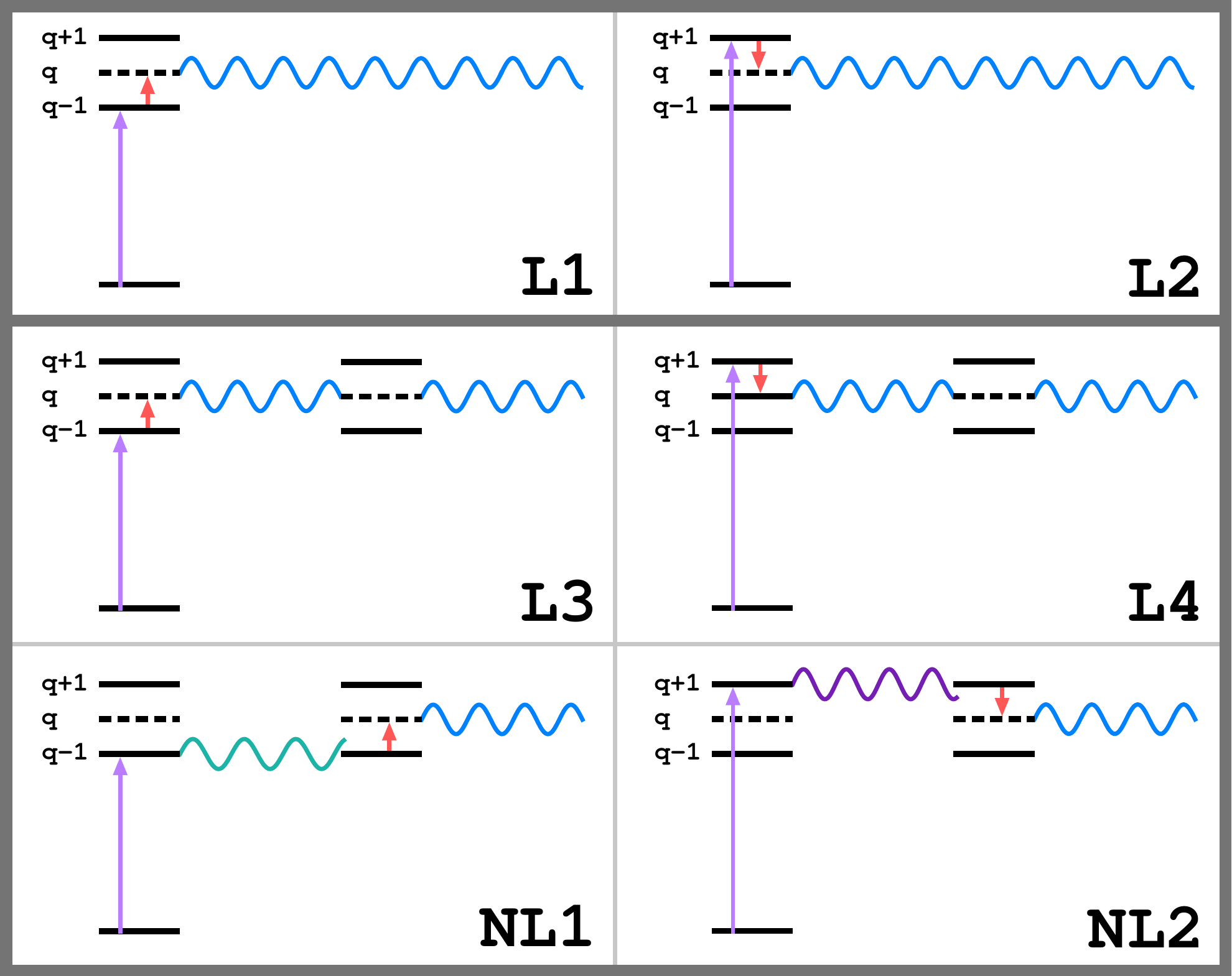}
      \caption{Schematic depiction of the four ``local'' (L1, L2, L3, L4) and the two ``non-local'' paths contributing mainly to the attosecond interference signal if one scattering event is possible after ionization.}
      \label{fig:paths}
    \end{figure}
    
%    The origin and the oscillation of $\tau_{\rm nl}$ can be understood from a multi-path model.
    
    Figure \ref{fig:paths} shows the relevant paths that an electron can take so that it ends up in the sideband $q$.
    The paths L1 and L2 are ``local'' paths that are the basis of APES in gas-phase molecules, the paths L3 and L4 are paths where interaction with the assisting IR field is still at the ionization site but the outgoing electron wave packet undergoes scattering, and NL1 and NL4 are ``non-local'' paths where photon exchange happens at the scattering center.
    In words, those paths are:
    \begin{itemize}
     \item L1: Absorption of an XUV photon of frequency $(q-1) \omega$ and absorption of one IR photon at the ionization site,
     \item L2: absorption of an XUV photon of frequency $(q+1) \omega$ and emission of one IR photon at the ionization site,
     \item L3: absorption of an XUV photon of frequency $(q-1) \omega$ and absorption of one IR photon at the ionization site, and scattering at the perturber without photon exchange,
     \item L4: absorption of an XUV photon of frequency $(q+1) \omega$ and emission of one IR photon at the ionization site, and scattering at the perturber without photon exchange,
     \item NL1: absorption of an XUV photon of frequency $(q-1) \omega$ at the ionization site and absorption of one IR photon at the perturber, and
     \item NL2: Absorption of an XUV photon of frequency $(q+1) \omega$ at the ionization site and emission of one IR photon at the perturber.
    \end{itemize}
    While L1 and L2 are the two paths that explain APES for gas-phase molecules, all six paths are needed reproduce the delay of Fig.\ \ref{model}.
    Each of these paths contributes to the second-order amplitude (i.e., the amplitude for two-photon transitions) $f$, i.e., 
    \begin{align}
      f = f_{\rm L1} + f_{\rm L2} + f_{\rm L3} + f_{\rm L4} + f_{\rm NL1} + f_{\rm NL2},
    \end{align}
    and the sideband intensity is then $I \propto |f|^2$.
    The amplitudes are
    \begin{align}
      f_{\rm L1} &= F_q^{q-1} e^{i k_q x_0} \\
      f_{\rm L2} &= F_q^{q+1} e^{i k_q x_0} \\
      f_{\rm L3} &= F_q^{q-1} e^{i k_q x_0} f_q^q \\
      f_{\rm L4} &= F_q^{q+1} e^{i k_q x_0} f_q^q \\
      f_{\rm NL1} &= F_{q-1}^{q-1} e^{i k_{q-1} x_0} f_q^{q-1} \\
      f_{\rm NL2} &= F_{q+1}^{q+1} e^{i k_{q+1} x_0} f_q^{q+1}.
    \end{align}
    Here, $F_n^m$ indicates the amplitude of transition from state $m$ to state $n$ at the photoionization site and $f_n^m$ indicates the amplitude of transition from state $m$ to state $n$ at the perturber.
    An additional phase factor is due to propagation to the scattering site with momentum $k_q$ (for L1, L2, L3, and L4), with momentum $k_{q-1}$ (NL1), or with momentum $k_{q+1}$ (NL2).
%     For photon absorption/emission, an additional phase factor $\mp \delta$ is added with $\delta = \omega \Delta t$ being the phase shift due to the offset between XUV and IR pulse.
    
    Numerical values for the amplitudes $F_n^m$ and $f_n^m$ can be obtained within what is known as strong-field approximation, soft-photon approximation, or Kroll-Watson theory, and which is referred to as laser-assisted photoelectric effect (LAPE) and laser-assisted electron scattering (LAES), respectively \cite{kroll73a,madsen05,galan13a}.
    For LAPE, it is assumed that the initial state is a bound state, the final state is that of a laser-dressed free electron in the IR field, the ionizing XUV pulse can be treated as perturbation, and the IR field is approximately a continuous wave.
    Similar approximations are made for LAES, with the initial and final states being laser-dressed states of the free electron in the IR field and the scattering potential is the perturbation.
    These theories provide the amplitudes 
    \begin{align}
      F_q^{q+n} &= e^{i n (\frac{\pi}{2} + \omega \Delta t)} J_n\left( \frac{k_q F_0^{\rm IR}}{\omega^2} \right) f_{q+n}^{\rm PI} \label{eq:lape1d} \\
      f_q^{q+n} &= e^{i n (\frac{\pi}{2} + \omega \Delta t)} J_n\left( \frac{(k_q-k_{q-n}) F_0^{\rm IR}}{\omega^2} \right) f_{q+n,q}^{\rm ES} \label{eq:laes1d}
    \end{align}
    where $F_0^{\rm IR}$ is the maximum field strength of the IR pulse, $f_{q+n}^{\rm PI}$ is the field-free photoionization amplitude, and $f_{q+n,q}^{\rm ES}$ is the field-free elastic scattering amplitude.
    $J_n$ are the Bessel function of the first kind.

    The side band intensity in this multi-path model is calculated as
    \begin{align}
      I \propto |f|^2 = A \cos\left(2 \omega ( \Delta t + \tau_{\rm PI} + \tau_{\rm nl}) \right), 
    \end{align}
    where the photoionization delay is given by the phase differences of the LAPE amplitudes,
    \begin{align}
      \tau_{\rm PI} = \frac{\arg(F_q^{q+1}) - \arg(F_q^{q-1})}{2 \omega}.
      \label{eq:taupi}
    \end{align}
    With some algebra, the non-local delay can be approximated as
    \begin{align}
      \tau_{\rm nl} 
        &\approx \frac{1}{\omega} \frac{A_{\rm nl}}{A_{\rm l}} 
          \sin\left( \frac{1}{2} \left( k^+ x_{\rm s}  + \phi^+ \right)\right) 
          \times 
          \sin\left( \frac{1}{2} \left( k^- x_{\rm s}  + \phi^- \right) \right),
          \label{eq:taunl}
    \end{align}
    where $A_{\rm l}$ and $A_{\rm nl}$ correspond to the (approximately equal) amplitudes of the processes where absorption or emission of the IR photon happens during photoionization and during scattering, respectively, i.e.,
    \begin{align}
      A_{\rm l}  &\approx |f_{\rm L1} + f_{\rm L3}| \approx |f_{\rm L2} + f_{\rm L4}| \\
      A_{\rm nl} &\approx |f_{\rm NL1}| \approx |f_{\rm NL2}|,
    \end{align}
    and the phase shifts of the two sine functions are\footnote{
      The $1$ in $1+f_q^q$ comes from combination of the paths L1 and L3 or L2 and L4.}
    \begin{align}
      \phi^+ &= \arg(f_q^{q+1}) - \arg(f_q^{q-1}) \\
      \phi^- &= \arg(f_q^{q+1}) - 2 \arg(1+f_q^q) + \arg(f_q^{q-1}).
    \end{align}
    The period of oscillation with $x_{\rm s}$ is characterized by the two wave numbers
    \begin{align}
      k^+ &= k_{q+1} - k_{q-1} \\
      k^- &= k_{q+1}-2 k_{q}+k_{q-1}.
    \end{align}
    Clearly, the non-local delay \eqref{eq:taunl} exhibits a beat pattern, i.e., an interference of two waves with slightly different oscillation frequencies. 
    In this case, the two frequencies giving rise to the beat pattern are $k_{q+1}-k_q$ and $k_q-k_{q-1}$, whose sum is $k^+$ (the fast oscillation with $x_{\rm s}$) and whose difference is $k^-$ (the slow oscillation with $x_{\rm s}$).
    We also note due to the phase shifts the non-local delay starts close to its maximum and the oscillation is cosine-like.

The first step in translating these results from a one-dimensional model system to the condensed phase consists of including the effect of path-length distributions. The distance travelled by an electron between photoionization and the first collision or between two collisions is, in general, a distributed quantity, the average of which is defined as the mean-free path. For simplicity, we first consider only one scattering event and we assume an exponential distribution of path lengths, which is the standard assumption in condensed-phase electron-transport simulations (see, e.g. Ref. \cite{nikjoo16a}). 

\begin{figure}[h!]
\begin{center}
\includegraphics[width=0.55\textwidth]{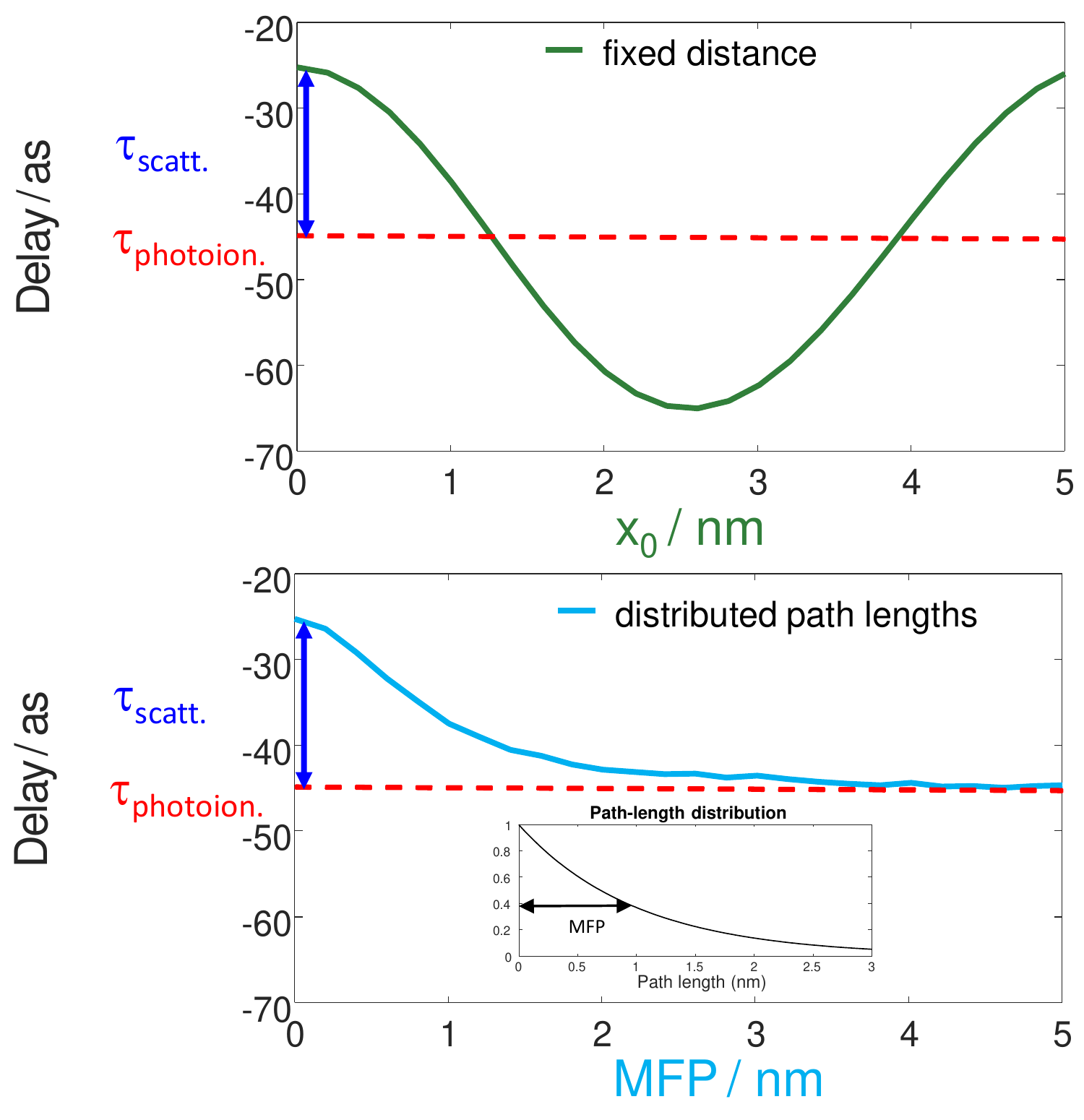}
\end{center}
\caption{{\bf Effect of path-length distribution}. The top panel illustrates the case of a fixed distance $x_0$ between photoionization and scattering, whereas the bottom panel represents the case where the path lengths follow an exponential distribution. In the bottom panel, the delay is given as a function of the average path length (MFP). }
\label{PLD}
\end{figure}

Figure \ref{PLD} shows the effect of the path-length distribution on the calculated delay. The top panel shows the results obtained when the distance between photoionization and scattering has a fixed value $x_0$, whereas the bottom panel shows the case where the distance is a distributed quantity following an exponential distribution, as a function of the average path-length MFP. We find that the oscillatory behavior from the fixed-distance case is turned into a monotonic decay from $\tau_{\rm photoion}+\tau_{\rm scatt}$ to $\tau_{\rm photoion}$. This result has an intuitive explanation. When the MFP is long, specifically MFP$\geq L$, where $L$ was defined above, the contribution of scattering cancels because of the oscillatory nature of the delay dependence on the path length. When MFP$\ll L$ the delays behave additively, i.e. the classical situation is recovered.

\begin{figure}[h!]
\begin{center}
\includegraphics[width=0.8\textwidth]{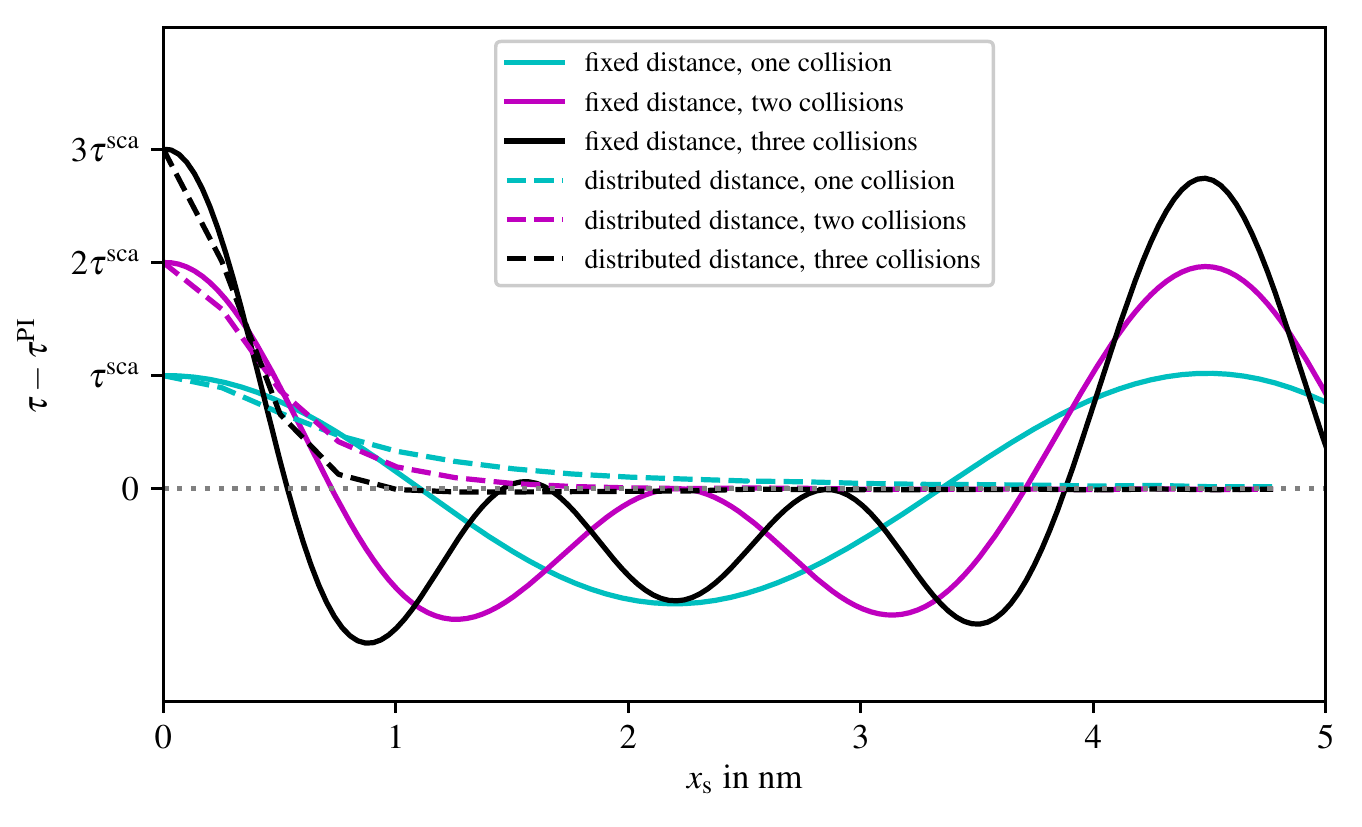}
\end{center}
\caption{{\bf Effect of multiple collisions}. Calculated delay for the case of a finite number of collisions separated by fixed distances $x_0$ (full lines) or path lengths that follow an exponential distribution of average $x_0$ (dashed lines).}
\label{MulColl}
\end{figure}

In a second step, we now include the treatment of multiple collisions. Figure \ref{MulColl} shows the results obtained for $n=1-4$ collisions. In the case of fixed distances (full lines), the total delay can expressed as
    \begin{align}
      \tau_{\rm nl} \approx \tau_{\rm sca} \sum_{j=1}^n \cos(n k x_{\rm s}),
    \end{align}
    where $\tau_{\rm scatt}$ is the scattering delay.
Thus, at $x_{\rm 0} = 0$, the non-local delay is ca.\ $\tau_{\rm photoion}+n \tau_{\rm scatt}$.
In the case of distributed path lengths (dashed lines)

These results are in line with the same intuitive explanation given above. For long MFPs the contribution of scattering vanishes, such that the calculated delay is equal to the photoionization delay. For very short MFPs ($\ll L$), the delays again behave additively, i.e. the classical limit is reached.

\subsubsection{Scattering delays and differential scattering cross sections of water}
\label{DCS}

The previous section has established the conceptual foundations of attosecond interferometry in condensed matter on the basis of simple model potentials. The purpose of this section is to present the results of state-of-the-art quantum scattering calculations on water clusters as a computationally tractable model of liquid water.

\begin{figure}[h!]
\begin{center}
\includegraphics[width=0.8\textwidth]{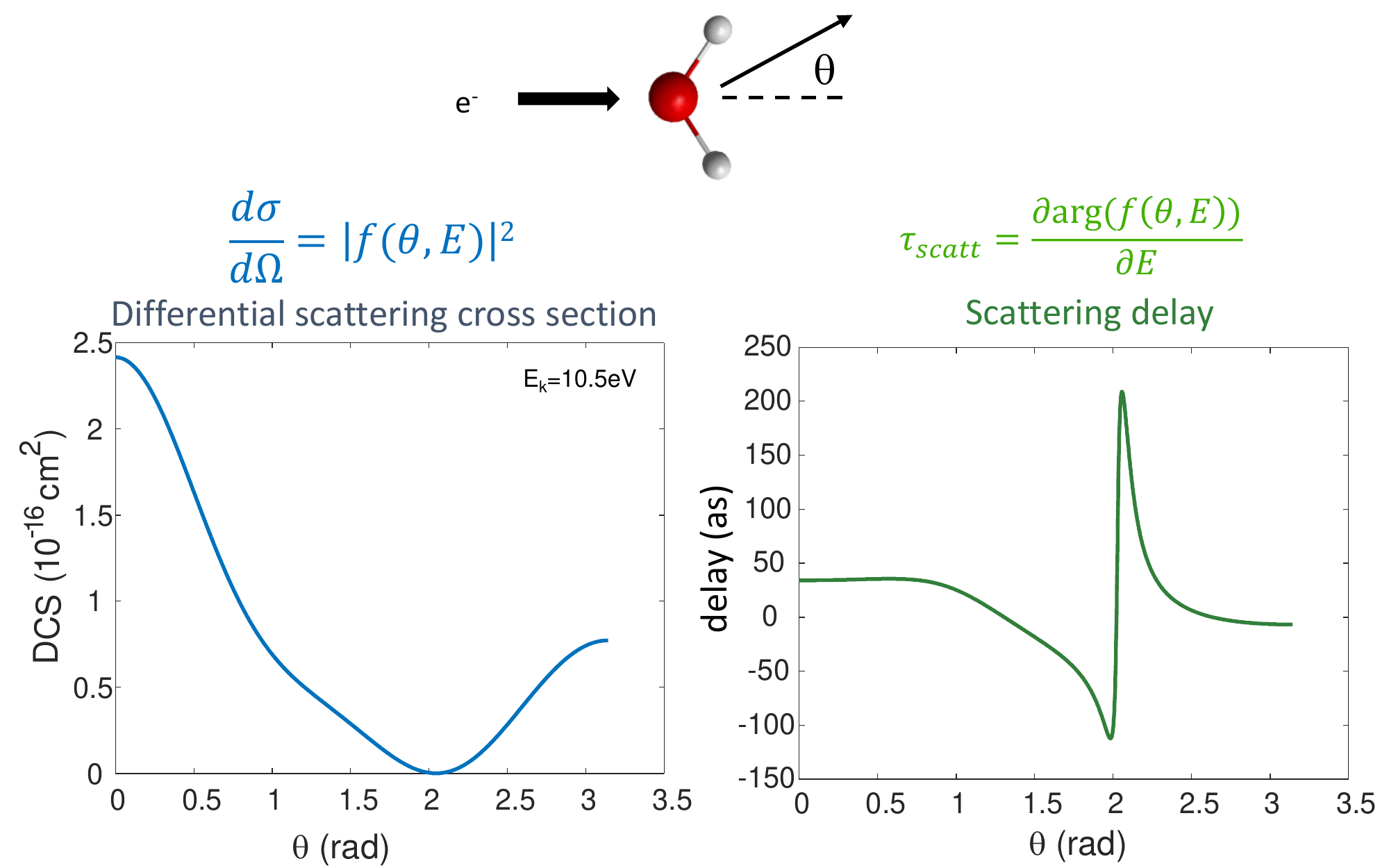}
\end{center}
\caption{{\bf Calculations of DCS and scattering delay for water monomer}. These calculations were done for an electron kinetic energy of 10~eV using the commercial R-matrix code Quantemol, in which the calculation of the DCS is implemented. The calculation of the scattering delay from the scattering factor provided by Quantemol was developed in house \cite{perry20b}.}
\label{monomer}
\end{figure}

Beginning with an isolated water molecule, we calculate the complex scattering factor for the fully elastic scattering process of a free electron with a water molecule in its rovibronic ground state. In our calculations, we use orientational averaging of the scattering target. As a consequence, the scattering problem has cylindrical symmetry, such that the scattering factor $f(\theta,E)$ depends only on the polar angle $\theta$ (see Fig. \ref{monomer}) and the electron kinetic energy $E$. This scattering factor defines both the differential scattering cross section (DCS) $\frac{d\sigma}{d\Omega}$ and the scattering delay $\tau_{\rm scatt}$ as defined in Fig. \ref{monomer}.

The DCS has a global maximum at $\theta=0$, corresponding to forward scattering and a local maximum at $\theta=\pi$, corresponding to back scattering. The scattering delay in the forward direction is positive, corresponding to the naive classical expectation. The scattering delay displays a rapid variation around $\theta\approx 2$~rad, i.e. at the same angle at which the DCS displays a local minimum. A detailed analysis of these calculations shows that the minimum in the DCS originates from a destructive interference of partial waves, which also leads to a rapid variation of the scattering delay.

\begin{figure}[h!]
\begin{center}
\includegraphics[width=0.8\textwidth]{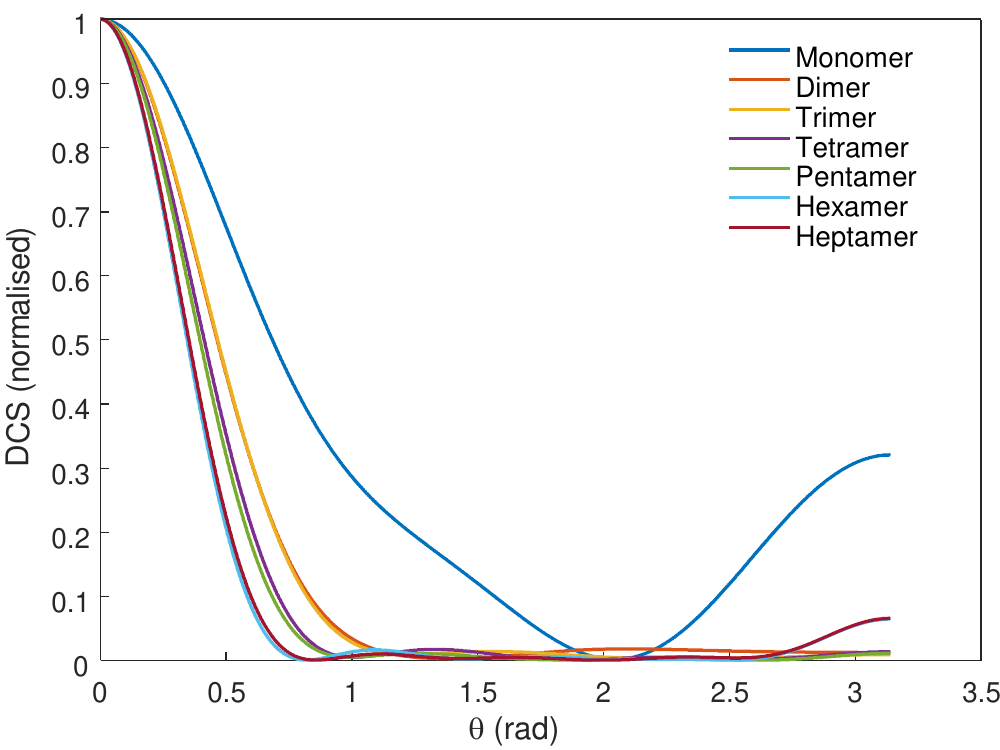}
\end{center}
\caption{{\bf Effect of condensation on the DCS of electron scattering}. Calculated DCS for electron scattering with water clusters of the indicated size using Quantemol. The equilibrium structures of the most stable isomers of the water clusters of each size have been taken from \cite{temelso11a}. All DCS have been normalized at $\theta =0$ for comparison.}
\label{clusters}
\end{figure}

Whereas these calculations are representative of electron scattering in the gas phase, and their quantitative accuracy has been verified \cite{perry20b}, they may not be representative of electron scattering in liquid water. Condensation is indeed known to modify the electronic structure of molecules, particularly in the case of water because of strong hydrogen bonding. We have therefore studied the evolution of the DCS with the size of the water cluster. The results are shown in Fig. \ref{clusters}.

The striking observation from these calculations is the rapid convergence of the DCS with increasing cluster size. The DCS evolves from a double-maximum structure with a relatively broad forward-scattering peak towards a quasi-Gaussian DCS without backscattering. The width of the quasi-Gaussian peak converges very rapidly with cluster size.

The rapid convergence of the DCS with the size of water clusters suggests that the highly accurate quantum-scattering calculations presented in this section can be used to model the description of electron scattering in liquid water. In addition to providing the crucial scattering factors required to model attosecond interferometry, they additionally enable us to determine both elastic and inelastic mean-free path for electron scattering in liquid water \cite{schild20a}, as discussed in the following section.

\subsubsection{Elastic and inelastic mean-free paths for electron scattering in liquid water}
\label{MFP}

\begin{figure}[htbp]
      \centering
      \includegraphics[width=0.99\textwidth]{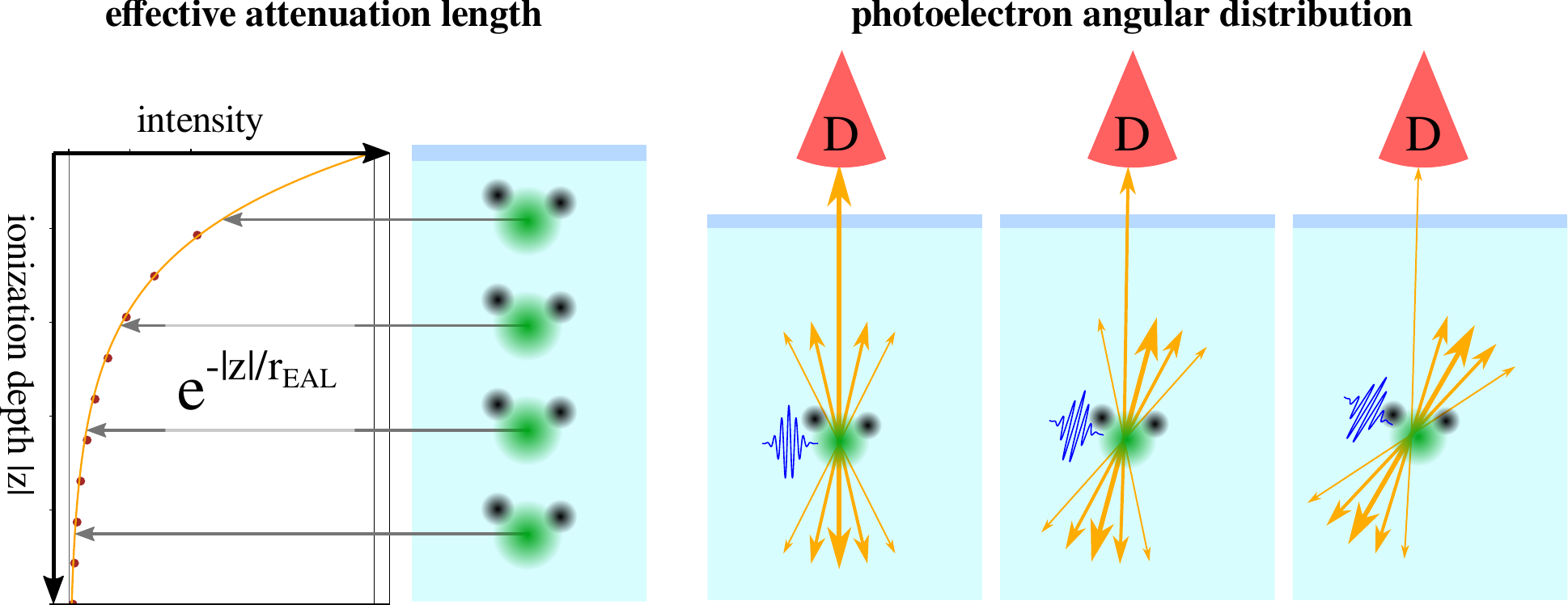}
      \caption{Left: Schematic depiction of the effective attenuation length (EAL) for ionization in a liquid. The signal of photoelectrons with kinetic energy corresponding to direct ionization decreases exponentially with the depth of the ionization site due to inelastic scattering. The EAL is the width parameter of the exponential function. Right: Idea of the measurement of the photoelectron angular distribution (PAD) of a liquid: Molecules are ionized with a beam and the photoelectron signal is measured with a detector above the surface that has a relatively small detection angle (indicated as ``D''). Variation of the beam polarization axis yields the PAD of the liquid.}
      \label{fig:padeal}
    \end{figure}

    Given the DCS for electron scattering at a certain kinetic energy, we are now in the position to determine the elastic and inelastic mean-free paths (EMFP and IMFP) \cite{schild20a}.
    For this purpose, we turn to two liquid-jet measurements of the photoionization of water:
    A measurement of the effective attenuation length (EAL) \cite{suzuki14a} and a measurement of the gas-phase and liquid-phase photoelectron angular distributions (PAD) \cite{thuermer13a} for oxygen 1s photoionization of water molecules.
    The underlying ideas of the experiments and of our simulations are depicted in Figure \ref{fig:padeal}.
    For ionization at a depth $|z|$ below the surface the measured photoelectron signal at the kinetic energy corresponding to direct photoionization behaves as 
    \begin{align}
      S(z) \propto e^{-|z|/r_{\rm EAL}},
      \label{eq:eal}
    \end{align}
    where $r_{\rm EAL}$ is the EAL.
    The loss of signal with $|z|$ is due to inelastic scattering, and the EAL is also a lower bound for the IMFP.
%    In \cite{suzuki14a}, $r_{\rm EAL}$ was determined by \dots
    We can simulate the experiment with our Monte Carlo trajectory code by using the field-free quantities and by simply measuring the number of electrons exiting the liquid depending on the assumed ionization depth.
    
    The PAD can, for achiral molecules, be described to a good approximation with an energy-dependent asymmetry parameter $\beta$ as
    \begin{align}
      \text{PAD}(\theta) \propto 1 + \beta P_2(\cos \theta),
      \label{eq:pad}
    \end{align}
    where $\theta$ is the polar angle and $P_2$ is the Legendre polynomial of second order.
    $\beta$-parameters were measured in \cite{thuermer13a} for the gas phase and for the liquid by rotating the polarization direction of the ionizing beam relative to a detector.
    In particular, the detector is not moved relative to the surface of the liquid jet, as indicated in Figure \ref{fig:padeal}.
    We can simulate also this experiment with our Monte Carlo trajectory code by rotating the PAD for ionization, which we take to be that of gas-phase water, as oxygen 1s-electrons are considered and we do not expect the corresponding orbitals to be influenced by solvation, except for a small shift of the binding energy. The detected signal is obtained by counting the electrons which exit the surface under a small polar angle.
    
    The program code for these simulations, CLstunfti, is publicly available \cite{clstunfti}.
    We indeed find the analytical behaviors given by \eqref{eq:eal} and \eqref{eq:pad} numerically.
    We also find the following dependencies of the EAL and the $\beta$-parameter $\beta_{\rm liq}$ of liquid water on the EMFP and on the average number of elastic scatterings $\langle N_{\rm ela} \rangle = \text{IMFP}/\text{EMFP}$:
    \begin{itemize}
      \item For fixed EMFP, the EAL increases linearly with $\sqrt{\langle N_{\rm ela} \rangle}$. This is because the EAL is essentially the root-mean-square translation distance of a random walk \cite{sethna06a}.
      \item For fixed $\langle N_{\rm ela} \rangle$, the EAL increases linearly with the EMFP. This happens because, for fixed $\langle N_{\rm ela} \rangle$, both the EMFP and the IMFP are scaled by the same amount and hence essentially all trajectories are also just scaled accordingly.
      \item For fixed $\langle N_{\rm ela} \rangle$, $\beta_{\rm liq}$ does not depend on the EMFP. This independence is also a result of the discussed scaling, as only trajectories are scaled but angles are invariant to changing both EMFP and IMFP by the same amount.
      \item For fixed EMFP, $\beta_{\rm liq}$ decreases to zero with increasing $\langle N_{\rm ela} \rangle$, i.e., the PAD becomes closer and closer to the isotropic PAD with each additional collision.
    \end{itemize}
    From these dependencies follows that a set of parameters $(r_{\rm EAL}, \beta_{\rm liq})$ corresponds to a unique set of mean free paths (EMFP, IMFP).
    We use this fact to find the EMFP and IMFP needed for our simulations.
    For sideband $14$ we obtain an EMFP of \unit[0.6]{nm} and $\langle N_{\rm ela} \rangle \approx 7$, and for sideband $20$ we obtain a slightly longer EMFP of \unit[0.8]{nm} and $\langle N_{\rm ela} \rangle \approx 6$.
    The values for the EMFP and IMFP from \unit[10]{eV} to \unit[300]{eV} determined with this approach can be found in \cite{schild20a}.

\subsubsection{Theoretical results for liquid water}

In this section, we combine the results described in the preceding subsections to a comprehensive model of APES in liquids. 
%This model includes, as its main novelty, the time delays originating from photoionization and scattering. It treats the interaction of the photoelectron wave packets with the laser fields on a quantum-mechanical and fully coherent basis. 
Starting from the discussion of the one-dimensional model system introduced in section \ref{NL1D}, we generalize these results to three dimensions and include the output from {\it ab-initio} quantum scattering calculations on realistic models of liquid water. For this purpose, we need the elastic and inelastic mean-free paths for electron transport in liquid water. These were determined based on two independent experiments on liquid-water microjets that we have inverted within our new formalism described in the previous section. The result is a fully self-consistent description of liquid-phase attosecond spectroscopy that incorporates the newly discovered non-local processes, as well as a state-of-the-art description of time delays in photoionization and scattering.

\begin{figure}[h!]
\begin{center}
\includegraphics[width=0.55\textwidth]{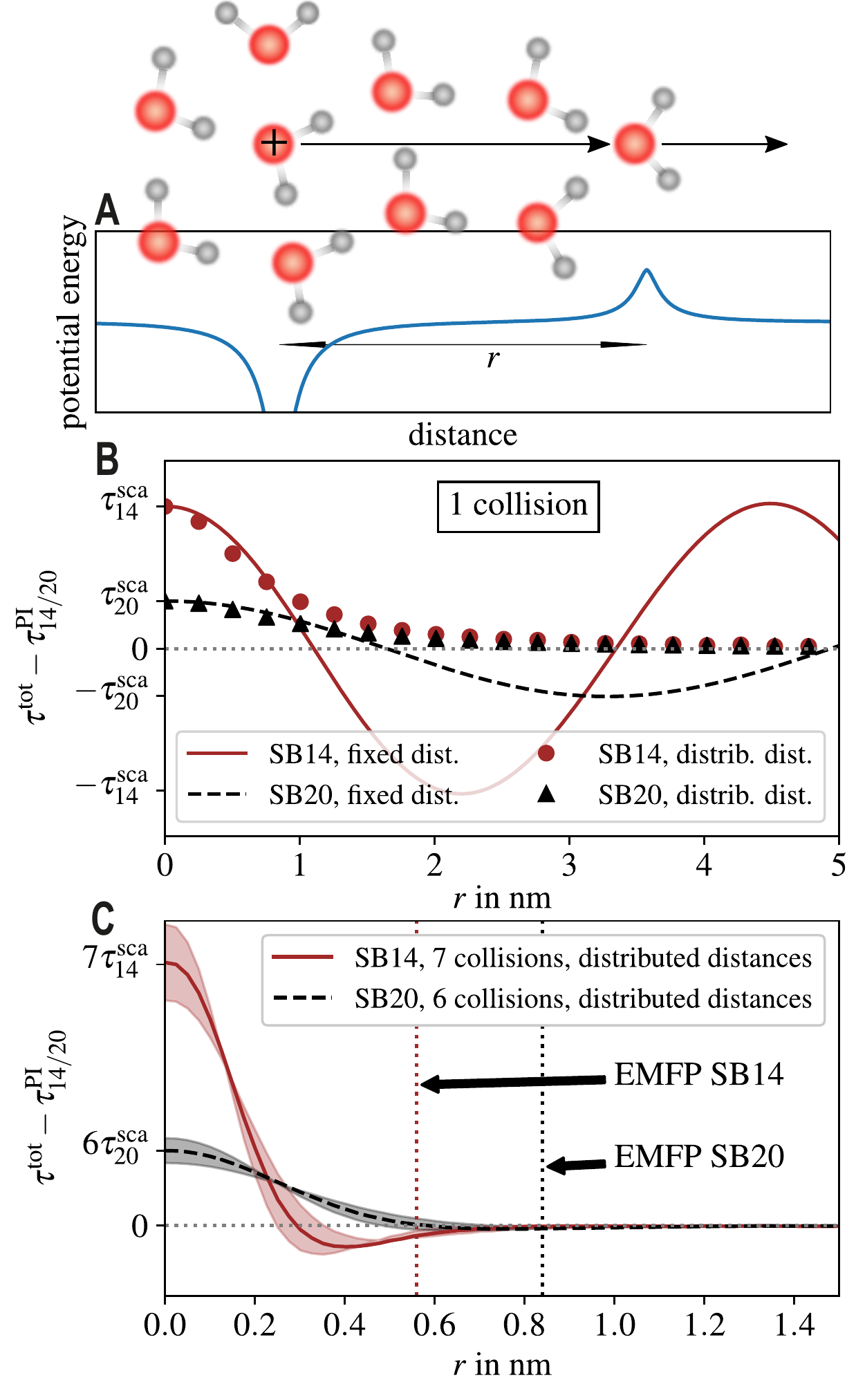}
\end{center}
\caption{{\bf Contributions of photoionization and scattering to the measured delays}. Schematic representation of the potentials used in the TDSE calculations (top) modelling the photoionization and scattering as attractive and repulsive potentials, respectively. The resulting delays ($\tau_{\rm tot}$) for the case of a single collision (middle) oscillate between $\tau_{\rm SB}^{\rm PI}+\tau_{\rm SB}^{\rm sca}$ and $\tau_{\rm SB}^{\rm PI}-\tau_{\rm SB}^{\rm sca}$ as a function of the distance $r$ (lines). In the case of an exponential path-length distribution with average $r$, a monotonic decay is obtained (symbols).
In the case of $n$ elastic collisions, sampled according to an exponential path-length distribution with average $r$, the delay decays from $\tau_{\rm SB}^{\rm PI}+n\tau_{\rm SB}^{\rm sca}$ to $\tau_{\rm SB}^{\rm PI}$ (bottom). Adapted from Ref. \cite{jordan20a}
%The relatively long EMFPs of liquid water, indicated as vertical lines, explain the cancellation of the scattering delays in the present results. This result was confirmed by the three-dimensional Monte-Carlo simulations described below.
}
\label{nonloc}
\end{figure}

As a first step towards interpreting the experimentally measured time delays, we return to the one-dimensional model system discussed in section \ref{NL1D}. The attractive potential is now chosen to mimic the ionized water molecule whereas the repulsive potential represents electron scattering with a neutral water molecule, as illustrated in Fig. \ref{nonloc}a. We additionally choose the photon energies to be those used in the experiment, i.e. harmonic orders 13/15 and 19/21 of an 800-nm driving field. Panel b shows the calculated delays as a function of the distance $r$ between the potentials representing photoionization and scattering. The full lines show the results of calculations done with fixed distances, whereas the symbol correspond to calculations assuming an exponential distribution of path lengths. These results confirm the conclusions of section \ref{NL1D}. Specifically, the total delay decays monotonically from $\tau^{\rm PI}_q+\tau^{\rm sca}_q$ to $\tau^{\rm PI}_q$ with increasing mean-free path. 

With the availability of the EMFP and IMFP for liquid water at the relevant electron-kinetic energies from section \ref{MFP}, we can now address the role of scattering delays in attosecond spectroscopy of liquid water within the one-dimensional model. The ratio IMFP/EMFP defines the number of elastic collisions that take place (on average) before one inelastic collision occurs. This ratio amounts to $\sim$7 in the case of SB14 and $\sim$6 in the case of SB20. Panel c therefore shows the delays obtained from one-dimensional simulations assuming these numbers of collisions, separated by distributed path lengths, as a function of the average path length (or EMFP) $r$. Although the variation of the calculated delay with $r$ is no longer strictly monotonic, as a consequence of the more rapid variations of the delay with $r$ in the case of multiple collisions (see Ref. \cite{rattenbacher18a} for details), the rapid convergence of both delays to $\tau^{\rm PI}_q$ is still obvious. Most importantly, when using the EMFP values determined in section \ref{MFP}, we find that both delays have practically converged to $\tau^{\rm PI}_q$. In other words, the one-dimensional calculations already suggest that the contributions of electron scattering during transport in liquid water are negligible under our experimental conditions. Although these results were obtained with simple model potentials, they already show that any possible contribution of scattering delays will be strongly reduced by the nature of the non-local processes. 

Before attempting an interpretation of the experimental results, we need to generalize this formalism to three dimensions and include a realistic description of photoionization and scattering in liquid water. Since the fully quantum-mechanical time-dependent description of electron transport in liquid water is computationally intractable, we opted for a hybrid quantum-classical approach. The analysis of the TDSE calculations in section \ref{NL1D} has shown that a quantum-mechanical description of photoionization and scattering within the Kroll-Watson formalism, combined with a classical-trajectory representation of electron transport, produces results of quantitative accuracy for electrons with kinetic energies $E_{\rm k}\gtrapprox$10~eV. Independent earlier work has shown that electron-transport simulations in liquid water based on a classical-trajectory Monte-Carlo description also reach quantitative agreement with a fully quantum simulation of electron transport, provided that the kinetic energy is larger than $\sim$10~eV \cite{liljequist08a}. We therefore based our description of liquid-phase attosecond spectroscopy on the same approach as section \ref{NL1D}, i.e. a quantum-mechanical description of photoionization, scattering and their laser-assisted counterparts and a classical-trajectory Monte-Carlo description of the electron transport steps.

This is done as follows:
    \begin{itemize}
      \item 
        We assume a flat interface between the liquid and the vacuum at $z=0$ (called the ``surface''), where $z<0$ is inside the liquid.
        Many trajectories are initialized at a randomly selected starting depth $z_0<0$.
        Those trajectories, taken together, represent the photoelectron wave packet originating from one ionization event, thus each trajectory has a phase.
      \item
        Half of these trajectories are initialized in state $n = q-1$ and the other half in state $n = q+1$, where a ``state'' corresponds to the kinetic energy $n \hbar \omega - E_{\rm b}$.
      \item 
        The generalization of formula \eqref{eq:lape1d} to three dimensions reads
        \begin{align}
          F_q^{q+n} &= e^{i n (\frac{\pi}{2} + \omega \Delta t)} J_n\left( \frac{\vec{k}_q \cdot \vec{F}_0^{\rm IR}}{\omega^2} \right) f_{q+n}^{\rm PI} \label{eq:lape3d},
        \end{align}
        which describes the amplitude of changing from state $q+n$ to state $q$.
        We know the photoelectron angular distribution $|f_{q+n}^{\rm PI}(\theta)|^2$ depending on the polar angle $\theta$ from photoionization calculations described in Section \ref{clusters} and we do not need the phase $\arg f_{q+n}^{\rm PI}(\theta)$, because we are only interested in the non-local contribution $\tau_{\rm nl}$ to the total measured delay $\tau = \tau_{\rm PI} + \tau_{\rm nl}$, cf.\ \eqref{eq:tausum} \footnote{\label{fn:angles}
          We use two sets of angles: The ``lab'' frame with polar angle $\theta$ and azimuthal angle $\phi$, where $\theta$ is the angle relative to the $z$-axis define via the surface normal, and the relative frame with polar angle $\vartheta$ and azimuthal angle (never explicitly used here), where $\vartheta$ is the angle relative to the direction of motion of the electron/trajectory. For the simulations described here the polarization of the laser pulses is assumed to be in $z$-direction, hence there is no difference between lab frame and relative frame for ionization. This is different for scattering.
          }
                
        Equation \eqref{eq:lape3d} can be used to model the interaction of the electron with the assisting IR field during photoionization.
        This is done by randomly selecting trajectories to change to states $q-2$, $q$, or $q+2$, according to the $|F_{q-2}^{q-1}|$ (emission), $|F_{q-1}^{q-1}|$ (no photon exchange), and $|F_{q}^{q-1}|$ (absorption) if the trajectory is initially in state $q-1$, and according to the amplitudes $|F_{q+2}^{q+1}|$ (absorption), $|F_{q+1}^{q+1}|$ (no photon exchange), and $|F_{q}^{q+1}|$ (emission) if the trajectory is initially in state $q+1$.\footnote{
          The probability to stay in the respective states is obtained from $|F_{q-1}^{q-1}|$ and $|F_{q+1}^{q+1}|$. For the relevant parameters, absorption or emission of two IR photons is irrelevant.}
        If a state change happens, the corresponding trajectories obtain a phase as given by \eqref{eq:lape3d} (without the unknown phase due to $f_{q+n}^{\rm PI}$).
        We note that as the laser field polarization is always assumed to be in $z$-direction, $F_q^{q+n}$ depends only on the offset $\Delta t$ between ionizing XUV and assisting IR pulse, the ionization direction, and the initial/final states.
      \item
        In the end, only trajectories reaching the surface in state $q$ contribute to the sideband of interest and hence are the only trajectories included in the determination of $\tau_{\rm nl}$.
        The probability of ending up in state $q$ is negligible if the trajectories end up once in $q \pm 2$, hence those trajectories are discarded.
      \item 
        We use mean-free paths $r_{\rm MFP}$ based on the general definition
        \begin{align}
          P(r) = \frac{1}{r_{\rm MFP}} e^{-r/r_{\rm MFP}}
          \label{eq:mfp}
        \end{align}
        to model random scattering in the medium.
        Each trajectory is attributed a randomly chosen maximal path to travel according to an IMFP $r_{\rm IMFP}$.
        This IMFP reflects (electronically) inelastic scattering such that the scattered electrons loose enough energy not to be detected at the kinetic energy of the considered sideband.
        If a trajectory reaches this maximum path length before reaching the surface, it is discarded.\footnote{
          There is another way to implement the IMFP which is computationally more efficient if many IMFP values should be tested: No maximum path length is set but the contributions of the trajectories reaching the surface in state $q$ to the total signal are weighted as $e^{-r_{\rm tot}/r_{\rm MFP}}$, where $r_{\rm tot}$ is the total path that the respective trajectory travelled until reaching the surface.}
      \item 
        The trajectories are moved in the direction chosen randomly according to $|f_{q+n}^{\rm PI}(\theta)|^2$ and for a distance $r$ chosen randomly according to \eqref{eq:mfp} with EMFP $r_{\rm EMFP}$.
        At this distance, elastic scattering is assumed to happen.
        A phase $k_n r$ is added to the phase of the trajectory, where $k_n$ is the momentum that the electron has in its current state.
      \item 
        The generalization of formula \eqref{eq:laes1d} for laser-assisted scattering to three dimensions is
        \begin{align}
          f_q^{q+n} &= e^{i n (\frac{\pi}{2} + \omega \Delta t)} J_n\left( \frac{(\vec{k}_q-\vec{k}_{q-n}) \cdot  \vec{F}_0^{\rm IR}}{\omega^2} \right) f_{q+n,q}^{\rm ES}
          \label{eq:laes3d}
        \end{align}
        where $f_{q+n,q}^{\rm ES}(\vartheta)$ is the field-free scattering amplitude depending on the polar angle $\vartheta$ relative to the current direction of motion, $\vec{k}_{q-n}$ is the momentum vector of the incoming electron, and $\vec{k}_q$ is the momentum vector of the outgoing electron.
        For each elastic scattering, a probability to change states as well as new directions for the motion of the electrons is chosen according to the amplitudes $f_q^{q+n}$.
        If a state change happens, the corresponding phase is added to the phase of the trajectory.
      \item 
        From $|f_q^{q+n}|^2$ new directions for the outgoing electron trajectories are determined and the process of elastic scattering is repeated.\footnote{
          We note that $f_q^{q+n}$ depends on three angles, for example on the polar angle of the incoming electron trajectory $\theta_{\rm in}$ and the polar and azimuthal angles $\theta_{\rm ou}, \phi_{\rm ou}$ of the outgoing electron trajectory in the lab frame, as well as on the offset $\Delta t$ and the initial and final state of the electron. If the laser field polarization axis was not aligned with the $z$-axis of the lab frame (defined by the surface normal), $f_q^{q+n}$ would also depend on the azimuthal angle $\phi_{\rm in}$ of the incoming electron trajectory.}
      \item 
        Trajectories are stopped if they end in states $q \pm 2$, if their path is longer than the previously sampled corresponding maximum path, or if they reach the surface at $z=0$.
      \item 
        Trajectories reaching the surface in state $q$ which originate from the same starting point, are added coherently. The contribution to the total second-order amplitude originating from a given point is calculated by coherent summation over all trajectories, which reach the detector with total acquired phases $\gamma_j$:
        \begin{align}
          f_k^{\rm w}(\Delta t) = \sum_j^{n_k^{\rm det}} f_{j,k}^{\rm t} = \sum_j^{n_k^{\rm det}} e^{\gamma_j(\Delta t,z_k^{\rm ini})}.
        \end{align}
      \item
        Simulations are run for a large number of initial positions $z_j^{\rm ini}$.
        As it is assumed that electrons originating from different initial positions $z_j^{\rm ini}$ in the liquid do not interfere, these contributions are summed incoherently,\footnote{
          There is, however, no significant change to the result if the summation is done coherently.}
        so that the total signal for a given value of the offset $\Delta t$ is
        \begin{align}
          \rho(\Delta t) = \sum_k \frac{n_k^{\rm det}}{n^{\rm tot}} |f_k^{\rm w}(\Delta t)|^2.
        \end{align}
      
    \end{itemize}

    \begin{figure}[htbp]
      \centering
      \includegraphics[width=0.99\textwidth]{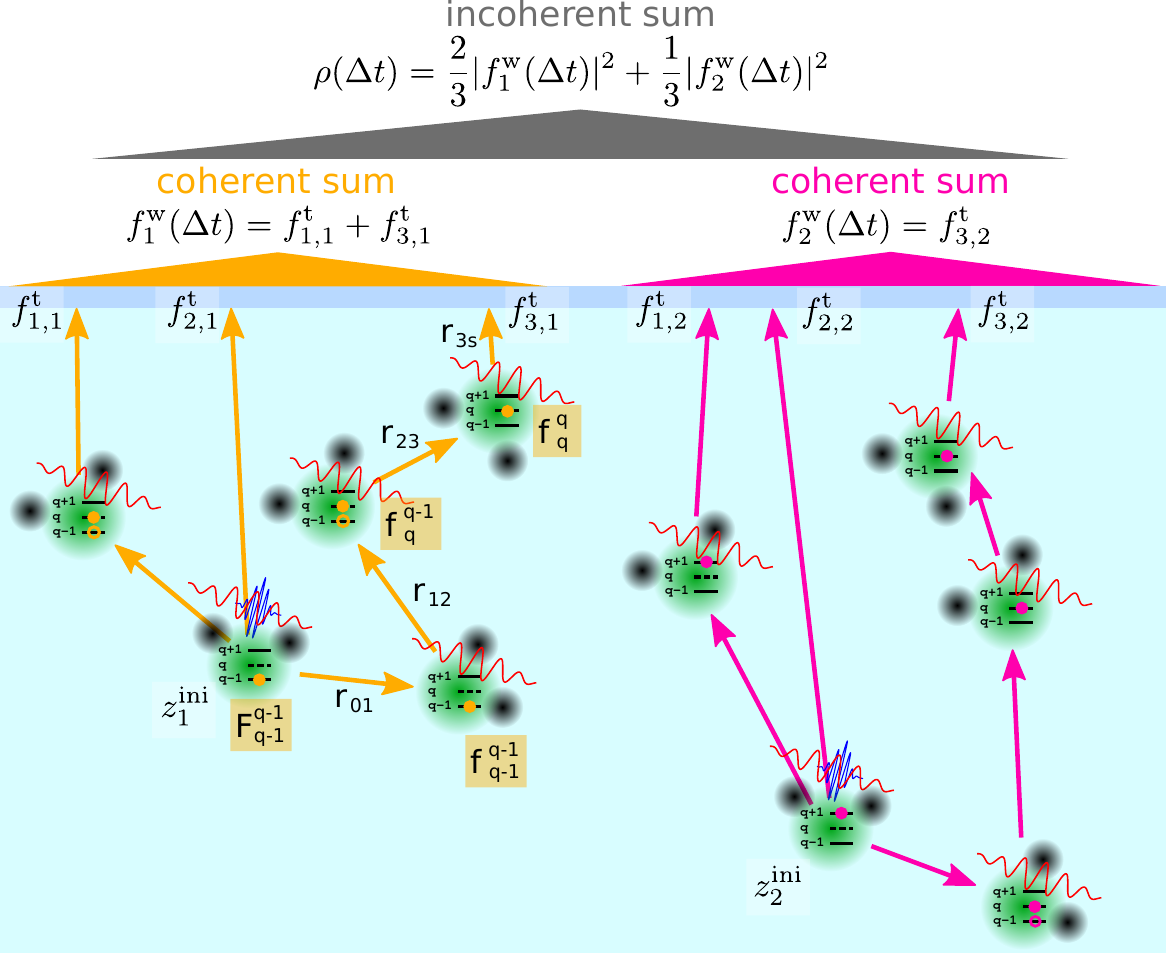}
      \caption{{\bf Three-dimensional model of attosecond interferometry in the condensed phase} Sketch of possible trajectories and how they are detected, i.e., how the intensity in the sideband $q$ is calculated as a function of the offset $\Delta t$ between XUV and IR pulse (the experimental control parameter). On the left side, three trajectories are shown which originate from a depth $z_1^{\rm ini}$, two of which end up in state $q$ ($f_{1,1}^{\rm t}$ and $f_{3,1}^{\rm t}$; for the latter, also path lengths $r_{jk}$ and amplitudes of the individual processes are given). The coherent sum of the corresponding amplitudes provides the contribution from $z_1^{\rm ini}$. On the right side, three trajectories originating from $z_2^{\rm ini}$, where only one trajectory ends in state $q$ ($f_{3,2}^{\rm t}$). Contributions for different $z_j^{\rm ini}$ are summed incoherently to obtain the sideband intensity $\rho(\Delta t)$.}
      \label{fig:3dscheme_sum}
    \end{figure}
    
    Figure \ref{fig:3dscheme_sum} illustrates the simulation procedure schematically.
    In the figure, two ionization sites (at $z_1^{\rm ini}$ and $z_2^{\rm ini}$) are depicted, from which three trajectories each are started.
    Let us first concentrate on one trajectory, i.e., the one with amplitude $f_{3,1}^{\rm t}$.
    This trajectory starts in state $q-1$, scatters once at a distance $r_{01}$ without photon exchange, scatters a second time at a distance $r_{12}$ and absorbs a photon to change to state $q$, and scatters one last time without photon exchange after a distance $r_{23}$ before it reaches the surface in state $q$ and is thus counted to the signal.
    The amplitude for the trajectory is thus 
    \begin{align}
      f_{3,1}^{\rm t} = F_{q-1}^{q-1} \, e^{i k_{q-1} r_{01}} \, 
                        f_{q-1}^{q-1} \, e^{i k_{q-1} r_{12}} \, 
                        f_{q-1}^{q  } \, e^{i k_{q  } r_{23}} \, 
                        f_{q  }^{q  } \, e^{i k_{q  } r_{3s}},
    \end{align}
    where the field-dressed ionization amplitude $F_{q-1}^{q-1}$ and all field-dressed scattering amplitudes depend on the directions along the path of the electron trajectory.
    The final signal obtained from the six trajectories is obtained as given in the figure.
    We note that the actual simulation uses millions to billions of trajectories for each ionization site, hundreds to thousands of ionization sites for convergence of $\tau_{\rm nl}$, and most of the trajectories do not end in state $q$ as the transition probabilities are relatively small.

Using the input parameters described in Sections \ref{DCS}-\ref{MFP}, we can simulate the contribution of the non-local delay to the total delay of APES for liquid water.
    We find that the non-local delay is almost independent of the average number of elastic collisions $\langle N_{\rm ela} \rangle \in [5,14]$ and decays with an increasing EMFP \unit[$\in$[0.1,1.3]]{nm}.
    A decrease of $\tau_{\rm nl}$ with increasing EMFP is expected from the one-dimensional model, as we see a decay of $\tau_{\rm nl}$ with the mean free path, cf.\ \ref{NL1D}.
    The independence of $\tau_{\rm nl}$ on $\langle N_{\rm ela} \rangle$ in the tested range is surprising but, as shown below, $\tau_{\rm nl}$ is also rather small.
    This independence may originate from a suppression of coherence by random scattering, or from the dominance of relatively short trajectories originating from close to the surface.
    
    \begin{figure}[htbp]
      \centering
      \includegraphics[width=0.99\textwidth]{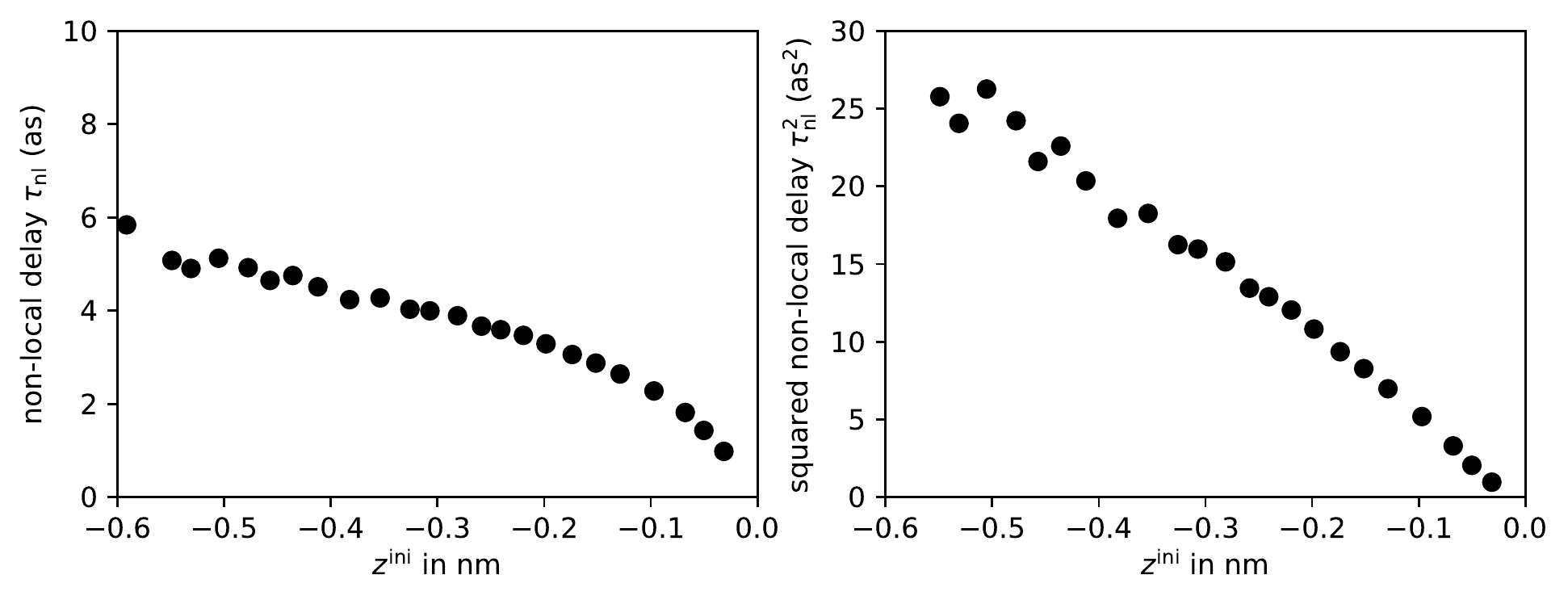}
      \caption{Dependence of the non-local delay $\tau_{\rm nl}$ for liquid water on the initial position $z^{\rm ini}$ of the ionized electron below the surface ($z=0$), as determined for the classical Monte Carlo trajectory model. This simulation was done for the parameters for sideband 14 (ca.\ \unit[10]{eV} electron kinetic energy) of an attosecond interferometry experiment. The right panel shows that $\tau_{\rm nl}(z^{\rm ini})^2$ is approximately a linear function. The deeper the electron starts, the less trajectories reach the surface, hence the accuracy of $\tau_{\rm nl}$ decreases with $|z^{\rm ini}|$ if the number of trajectories is not increased.}
      \label{fig:zdependence}
    \end{figure}
    
    Figure \ref{fig:zdependence} shows the dependence of the non-local delay (for EMFP=\unit[0.55]{nm} and $\langle N_{\rm ela} \rangle = 11$) on the initial position of the trajectories below the surface.
    We can see that $\tau_{\rm nl}$ becomes larger with starting depth, but is only ca.\ \unit[5]{as} for an initial depth of \unit[0.5]{nm}.
    As the contribution to the total signal also decays exponentially with starting depth, this graph already suggests that $\tau_{\rm nl}$ is small compared to the experimentally measured value of the delay.
    We find indeed that according to our simulations, the contribution of $\tau_{\rm nl}$ to the total delay is only about \unit[2]{as}.

        Thus, we conclude that the non-local delay is negligible in our attosecond experiments on liquid water.
    Nevertheless, there are situations where the non-local delay may play a role, for example if the EMFP is significantly shorter than in the present case or if the considered medium has long-range order, such as e.g. a crystal. For these cases, the non-local delays may become measurable, which would provide additional information regarding the scattering delays and mean-free paths.

\subsection{Interpretation}
\label{interpr}

We now return to the interpretation of the experimental results described in section \ref{exp}. These measurements have yielded time delays between photoemission from liquid and gaseous water of 69$\pm$20~as at 21.7~eV photon energy and 48$\pm$16~as at 31.0~eV photon energy. 
In addition, relative modulation contrasts of 0.17$\pm$0.03 and 0.45$\pm$0.06 have been determined at the mentioned photon energies. 
In section \ref{concepts}, we have introduced the concepts relevant to interpret these experimental results and quantitative methods to model such experiments. These calculations have shown that attosecond spectroscopy of liquids is, in general, sensitive to photoionization delays, scattering delays and mean-free paths. In the case of liquid water at the relatively low photon energies studied so far, the contributions of electron scattering and transport are predicted to be negligible. This leaves one main possible source for the measured delays, i.e. the photoionization step.

To evaluate the possible contribution from the photoionization step to the measured delay, we follow the same strategy as in sections \ref{DCS}-\ref{MFP}, i.e. we use accurate quantum-scattering calculations on water clusters as a model for liquid water. Our calculations are based on the formalism described in Ref. \cite{baykusheva17a}, which was developed to interpret the first measurements of photoionization delays in isolated molecules \cite{huppert16a}. Whereas the experimental measurements can at present only access relative time delays, e.g. between gaseous and liquid water, the calculations yield absolute time delays. We will therefore discuss the absolute photoionization time delays of water clusters and isolated water molecules. Finally, we will compare the relative time delays to the experimental results.

There are several reasons why photoionization time delays of water molecules can be expected to change under the effect of condensation. First, the valence electronic structure of water molecules changes with the addition of neighboring molecules. This is a consequence of the relatively strong dipole moment of the water molecule, which causes the formation of strong hydrogen bonds between water molecules. The consequence is a significant change in the electronic structure, which becomes visible in the hybridization of the molecular valence orbitals. In addition to the local change of the electronic wave function, condensation also causes a partial delocalization of the one-electron wavefunctions (orbitals) over more than one molecule. These effects concern the initial bound electronic wave function. They will all have an influence on the photoionization delay. Second, the final continuum wave function is also affected by condensation, probably even more than the initial state. The continuum wavefunction for bulk liquid water is very complex. It consists of a complicated conduction-band part inside the liquid bulk and a simpler part in the vacuum region outside the liquid. However, all that is needed to accurately describe photoionization is the continuum wave function over the spatial region where the initial-state wave function has a significant amplitude. Therefore, our calculations on water clusters can be expected to converge to the results for liquid water when the water clusters become sufficiently large as to describe the relevant spatial extension of the initial-state wave function. Both effects discussed in this paragraph are naturally included in our quantum-scattering calculations.

\begin{figure}[h!]
\begin{center}
\includegraphics[width=\textwidth]{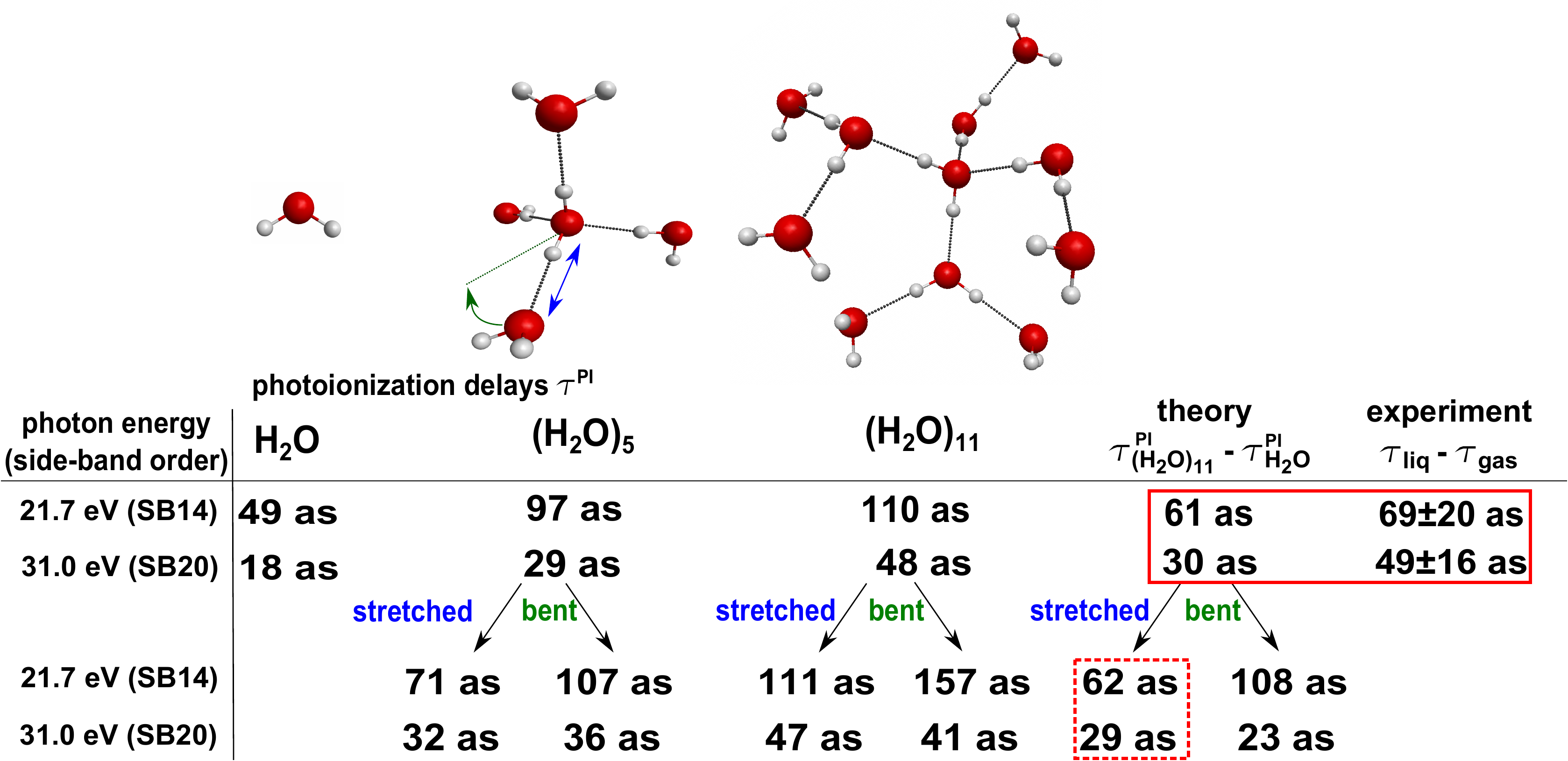}
\end{center}
\caption{{\bf Effect of solvation and hydrogen bonding on photoionization delays}. Calculated photoionization delays for H$_2$O, (H$_2$O)$_5$ and (H$_2$O)$_{11}$, representing water molecules with zero, one complete or one complete and one partial solvation shells, respectively. The bottom row indicates delays obtained by stretching (blue arrow) or bending (green arrow) one hydrogen bond in the clusters. Adapted from Ref. \cite{jordan20a}} 
\label{delays}
\end{figure}

Figure \ref{delays} shows the calculated photoionization delays of water clusters of increasing size at the two photon energies relevant for this work. These calculations were done with ePolyScat  \cite{gianturco94a,natalense99a} using the methods described in Ref. \cite{baykusheva17a}. The first two lines of the table show the photoionization delays of isolated water molecules, a water molecule with a complete first solvation shell (pentamer) and with a partial second solvation shell in which all dangling hydrogen bonds of the first shell have been coordinated (undecamer). Interestingly, the photoionization delays systematically increase with increasing size of the water cluster. Whereas the calculations at the lower photon energy show evidence of convergence with cluster size, the convergence is less obvious at the higher photon energy. Since converged quantum-scattering calculations on (H$_2$O)$_{11}$ reach the limit of current computational methods ($\sim$180 CPU days for one calculation), these calculations could not yet be extended to larger water clusters. However, the relative photoionization delays between the water clusters and the water monomer are in rewarding agreement with the experimental results. Whereas the calculated relative delay of 61~as at 21.7~eV agrees with the experiment within the error bar, the relative delay of 30~as at 31.0 eV is close to the error interval of the experiment. 

In addition to reaching near-quantitative agreement with the experiment, the results in the two bottom lines of Fig. \ref{delays} shed light on the other experimental observable introduced in this work: the modulation contrast in attosecond interferometry. Whereas the calculations shown in the upper two lines assumed a tetrahedral coordination geometry of each water molecule with O-O distances fixed to 2.8 \AA, the local solvation environment is known to be partially distorted in liquid water and subject to fluctuations on picosecond time scales. To simulate the effect of the structural distortions of the hydrogen bonds, we have stretched or bent one of the hydrogen bonds by 0.7 \AA~or 50$^{\circ}$, respectively, following the distortions used in Ref. \cite{wernet04a}. In the case of the pentamer, we have moved one of the water molecules, and in the case of the undecamer, we have moved a group of 3 water molecules while keeping its internal geometry unchanged. The effects of these local distortions of the hydrogen-bond structure is remarkable. The delays calculated at 21.7~eV display a pronounced sensitivity to the structure with variations reaching -16~as in the pentamer and +47~as in the undecamer. The corresponding changes at 31.0~eV are much smaller and do not exceed $\pm$7~as. These different sensitivities of the delays to the local solvation structure are consistent with the very different modulation contrasts observed at the two energies, i.e. 0.17$\pm$0.03 and 0.45$\pm$0.06, respectively. This provides an important first indication that the origin of the observed modulation depths might be the distribution of local solvation structures in liquid water. It does not exclude, of course, that other contributions, such as electron decoherence during transport through the liquid phase, also contribute to the finite modulation depths.

These results complete the consistent picture that we have built up along this book chapter. Whereas our calculations predicted that the contributions of electron scattering and transport were small, such that only the photoionization delays would play a role, the independent calculations of these photoionization delays on water clusters have confirmed this prediction, reaching a near-quantitative agreement with the experimental delays. The remarkable sensitivity of the delays to the local solvation structure offers interesting perspectives for extending such measurements to other photon energies, other liquids and solvated species. The investigation of local solvation structures around different types of solutes, e.g. anions vs. cations, are particularly promising, as well as the investigation of other attosecond time-scale processes in the liquid phase.

\section{High-harmonic spectroscopy of liquids}

High-harmonic generation (HHG) in gases has been extensively studied over the last three decades. Apart from its quintessential significance as a coherent light source, it has also been developed into a unique spectroscopic technique. High-harmonic spectra indeed contain a wealth of information about the structure and dynamics of the medium from which they are emitted. In the gas phase, HHG can be understood as a process that involves strong-field ionization, electron propagation in the continuum and photorecombination. As a consequence, high-harmonic spectra indeed contain information about the electronic structure of the medium which is encoded in the orientation-dependence of the strong-field-ionization and photorecombination dipole matrix elements. They also contain dynamical information as a consequence of the unique mapping from the transit time of the electron in the continuum to the emitted photon energy, when the contributions of the ''short'' electron trajectories are recorded.
The applications of high-harmonic spectroscopy (HHS) have led to many important results, such as the tomography of molecular orbitals \cite{itatani04a}, the observation of structural and electronic dynamics on attosecond time scales \cite{baker06a,smirnova09a}, the identification of two-center interference minima \cite{vozzi05a,vozzi11a}, the observation of Cooper minima \cite{woerner09a}, the determination of ionization and recombination times in HHG \cite{shafir12a} and the measurement and laser control of attosecond charge migration \cite{kraus15a}. The field of HHS has recently been reviewed in Ref. \cite{kraus18a}.

High-harmonic generation in solids is a much younger and highly dynamic research field. Although harmonic generation up to the 7$^{\rm th}$ order of a MIR driver in ZnSe has already been observed in 2001 \cite{chin01a}, the rapid development of HHS of solids has started with the observation of high-order harmonic generation in bulk solids in 2011 \cite{ghimire11a}, the analysis of inter- and intraband contributions to HHG \cite{vampa14a,vampa15a,luu16a} and the observation of extreme-ultraviolet HHG from solids \cite{luu15a}. HHS of solids has developed very rapidly, including impressive developments with terahertz drivers \cite{schubert14a,hohenleutner15a}, observations of crystal-structure effects \cite{you17a}, the measurement of the Berry curvature of solids \cite{luu18b} and the prediction of the possibility to observe phase transitions in solids \cite{ayuso19a}. The field of solid-state HHS has recently been reviewed in Ref. \cite{ghimire19a}.

In remarkable contrast to these developments, high-harmonic generation from liquids has barely been developed so far. Although relatively low-order harmonic generation from liquids in the visible domain has been observed in 2009 \cite{dichiara09a}, it took until 2018 before HHG from bulk liquids was reported \cite{luu18a}. This situation is certainly to some extent the consequence of the technological challenges associated with HHG from liquids. Early efforts to observe liquid-phase HHG indeed date back to 2003, when HHG from liquid-water droplets was attempted \cite{flettner03a}. Instead of observing the expected coherent HHG emission, the authors only observed incoherent emission from a plasma formed by multiple ionization of water molecules. Coherent XUV emission was only observed following the action of a pump pulse preceding the HHG pulse that caused a hydrodynamic expansion of the droplets \cite{flettner03a}. These results were confirmed in later experiments and combined with a model of the droplet expansion \cite{kurz13a,kurz16a}, which led to the conclusion that HHG was not observed at the density of liquid water but only appeared at densities that were significantly lower. The results reported in Ref. \cite{luu18a} contrast with this interpretation by showing that coherent HHG does occur at the density of liquid water. Finally, high-harmonic emission from a plasma created through the interaction of a laser with a liquid microjet has been observed in Ref. \cite{heissler14a}. These results were obtained in the so-called coherent-wake-emission regime, which is reached for extremely-high laser intensities ($I >$10$^{18}$W/cm$^{2}$). In this regime, HHG takes place from laser-induced plasma dynamics, which are insensitive to the properties of the target used to create the plasma. In this section, we discuss the methods, results and interpretations of HHG from bulk liquids, which form the basis of liquid-phase HHS, and introduce a new type of targets for strong-field science and the development of new high-harmonic light sources.

\subsection{Experimental methods}

\begin{figure}[h!]
\begin{center}
\includegraphics[width=\textwidth]{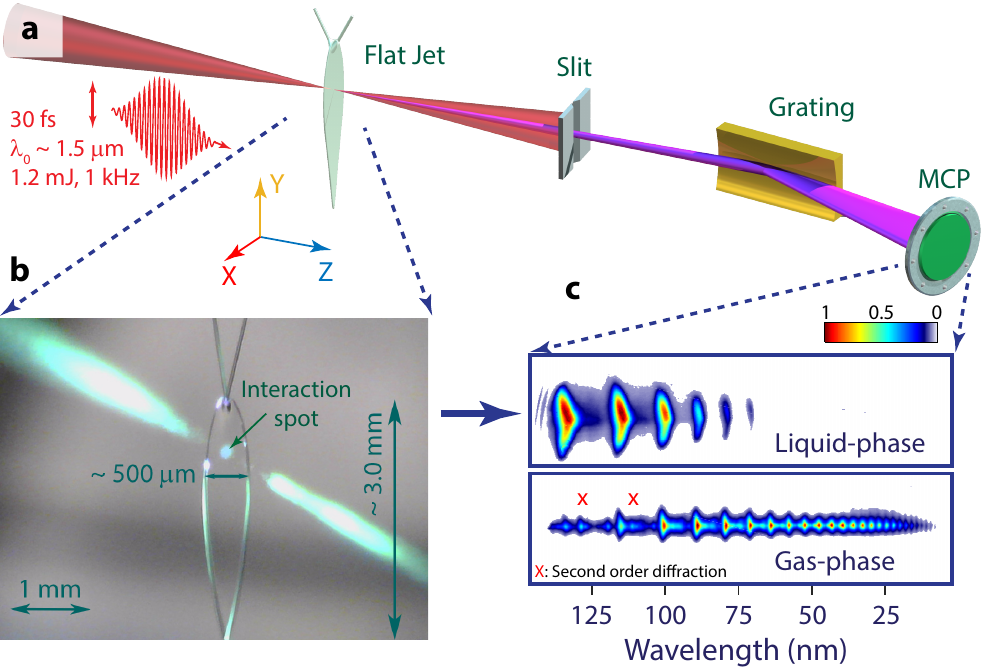}
\end{center}
\caption{{\bf Observation of HHG from liquids}. a) schematic representation of the experimental setup without the vacuum chambers, b) photograph of the flatjet under operating conditions, c) simultaneously recorded high-harmonic spectra of liquid and gaseous ethanol. Adapted from Ref. \cite{luu18a}.} 
\label{LHHG-setup}
\end{figure}

The first successful observation of extreme-ultraviolet HHG from bulk liquids was realized with the experimental setup illustrated in Fig. \ref{LHHG-setup} \cite{luu18a}.
The key enabling technology for this experiment was the flatjet technique \cite{hasson64a,ekimova15a}. The flatjet is created by colliding two cylindrical microjets under an impact angle of 48$^{\circ}$. In the present experiment, cylindrical jets with a diameter of 50 $\mu$m were used, resulting in a flatjet with a thickness of $\sim$1-2~$\mu$m. Since the thickness of the flatjet scales quadratically with the diameter of the colliding jets under reasonable assumptions \cite{hasson64a}, the thickness can easily be reduced, which however comes at the cost of a smaller cross section of the flatjet.
High-harmonic generation is realized by focusing a short-wave infrared (SWIR) femtosecond pulse centered at 1.5 $\mu$m onto the flatjet in normal incidence, as shown in Fig. \ref{LHHG-setup}a. The high-harmonic emission is detected by a flat-field spectrometer consisting of an entrance slit, a concave variable-line-spacing grating and a microchannel-plate detector backed with a phosphor screen and a charge-coupled device camera. A photograph of the flatjet under operating HHG conditions is shown in Fig. \ref{LHHG-setup}b. The green light originates from scattering of the third-harmonic radiation. Figure \ref{LHHG-setup}c shows the simultaneously observed high-harmonic spectra emitted from bulk liquid ethanol and the surrounding gas-phase ethanol. The two spectra were independently normalized to their maximal intensity. The comparison of the two spectra immediately reveals several characteristic differences, i.e. the liquid-phase harmonics have i) a much lower cut-off, ii) a larger divergence and iii) a rapidly decreasing intensity distribution compared to the gas-phase harmonics. 

\begin{figure}[h!]
\begin{center}
\includegraphics[width=\textwidth]{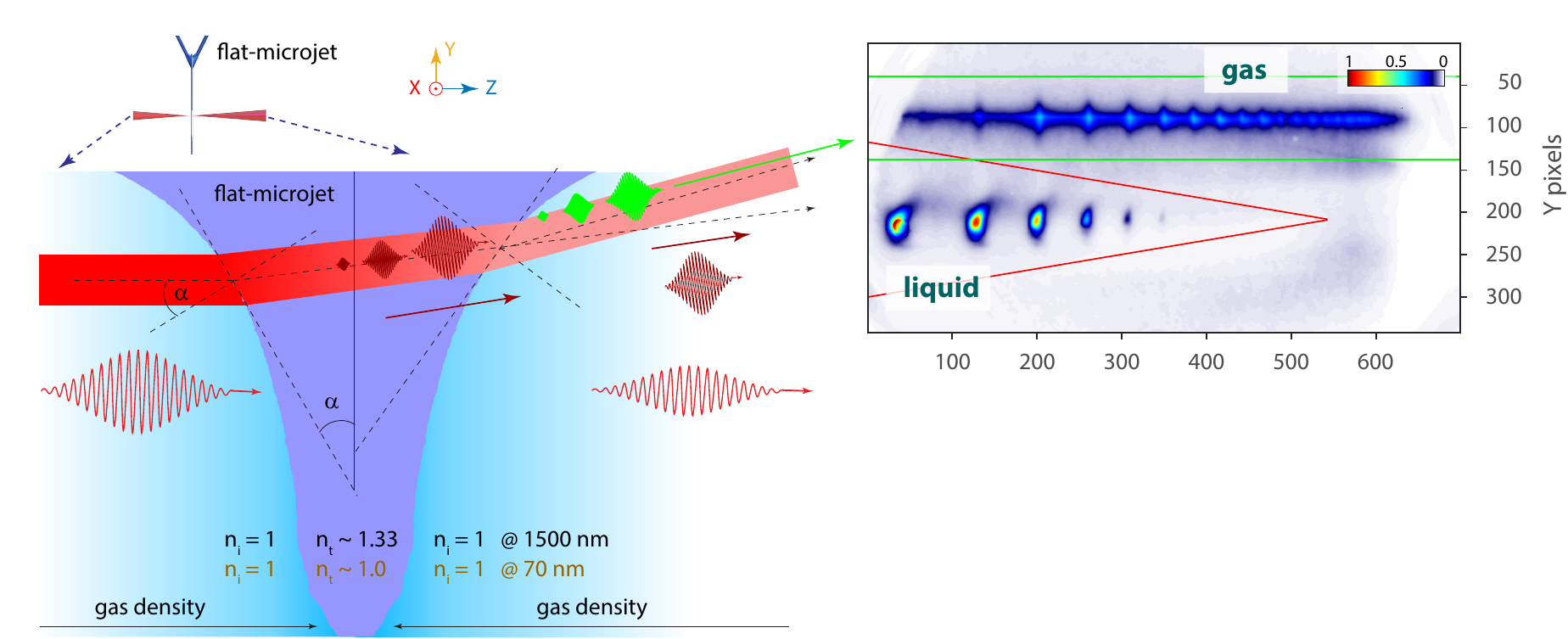}
\end{center}
\caption{{\bf Separation of HHG from the liquid and gas phases}. A schematic representation of the optical path of the fundamental beam across the liquid jet is shown on the left. The curvature of the jet surfaces has been exaggerated for clarity. A spatio-spectrally resolved far-field image of the high-harmonic spectra simultaneously emitted from the liquid and gas phases is shown on the right. Adapted from Ref. \cite{luu18a}.} 
\label{LHHG-sep}
\end{figure}

The most important aspect of liquid-phase HHG is the separation of the signals from the liquid and gas phases. The operation of a liquid jet in vacuum necessarily entails the presence of a surrounding gas phase created by evaporation from the liquid jet. The challenge of separating HHG from the gas and liquid phases is elegantly solved by the natural shape of the flatjet, as illustrated in Fig. \ref{LHHG-sep}. Since the thickness of the flatjet decreases from top to bottom across the first sheet, the interfaces are not parallel to each other but form a small angle with respect to each other. This leads to two consecutive refractions of the fundamental driving field upon entrance and exit of the liquid phase. The XUV radiation generated within the liquid jet is not (significantly) refracted when it exits the liquid because the index of refraction change is negligible in the XUV. This leads to a natural separation of liquid- and gas-phase HHG on the detector. The observed gas-phase HHG originates entirely from the gas phase located behind the flatjet because the high-harmonic radiation created in front of the jet is completely absorbed, given the typical absorption lengths of $\sim$10~nm at 20~eV \cite{hayashi15a}. This natural separation of HHG has several immediate benefits, which include the possibility to perform HHS experiments on the gas and liquid phases simultaneously and the separation of generated high harmonics from the driving fields, which could become relevant for high-average-power attosecond \cite{hammerland19a} or XUV-frequency-comb \cite{corder18a} experiments.

\begin{figure}[h!]
\begin{center}
\includegraphics[width=0.7\textwidth]{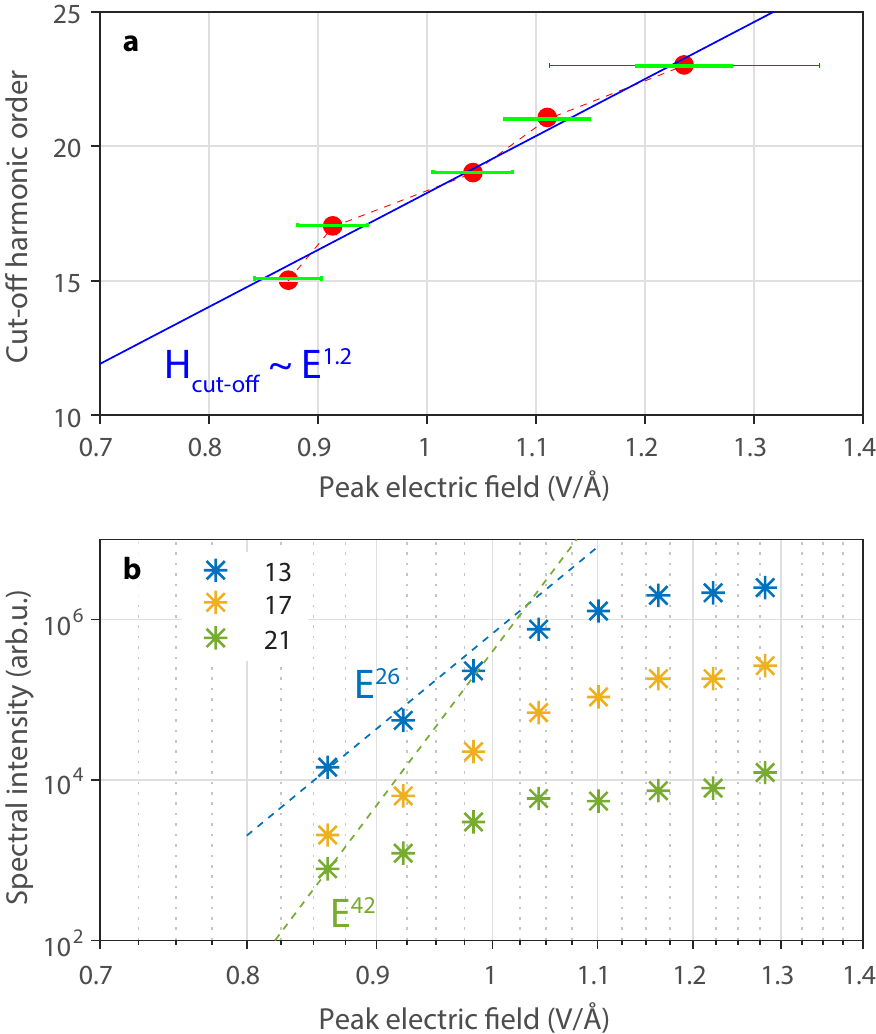}
\end{center}
\caption{{\bf Scaling of the liquid-phase HHG cutoff and harmonic yields}. Scaling of the cut-off harmonic order (a) and the yield of selected harmonic orders (b) with the peak electric field of the driver ($\sim$30~fs, centered at 1.5~$\mu$m) in liquid ethanol. Adapted from Ref. \cite{luu18a}.} 
\label{LHHG-scaling}
\end{figure}

One of the most fundamental characteristics of HHG is the scaling of its cut-off energy with the intensity of the driving field. In gases, the cut-off scales quadratically with the peak electric field of the driver. This can intuitively be understood as the signature of the maximal kinetic energy that the continuum electron can acquire from the driving field. The experimental results obtained on liquid-phase HHG are shown in Fig. \ref{LHHG-scaling}. They reveal a quasi-linear scaling ($H_{\rm cut-off}\propto E^{1.2}$) of the cut-off photon energy with the peak electric field of the driver. This result points at a fundamentally different mechanism of HHG in liquids compared to gases. The obtained result is similar to observations made in solids, where a quasi-linear scaling of the cut-off energy was also observed \cite{ghimire11a}. This cut-off scaling is common to both types of mechanisms discussed for solids, i.e. intra-band currents (Bloch oscillations) and interband polarization (generalized recollision) \cite{luu15a,vampa15a,luu16a}.

We next turn to the scaling of the high-harmonic yield with the driving-field intensity. Figure \ref{LHHG-scaling} shows the dependence of the yield of H13 and H21, generated from a 1.5~$\mu$m driver in liquid ethanol on the peak electric-field strength. An electric field strength of 1~V/\AA~corresponds to a peak intensity of 2.65$\times 10^{13}$~W/cm$^2$. The dashed lines of the same color indicate the corresponding perturbative scaling laws. All observed harmonic orders thus follow non-perturbative scaling laws with a deviation from the perturbative scaling that strongly increases with harmonic order. These results demonstrate the strongly non-perturbative character of the observed HHG from liquids.

\begin{figure}[h!]
\begin{center}
\includegraphics[width=\textwidth]{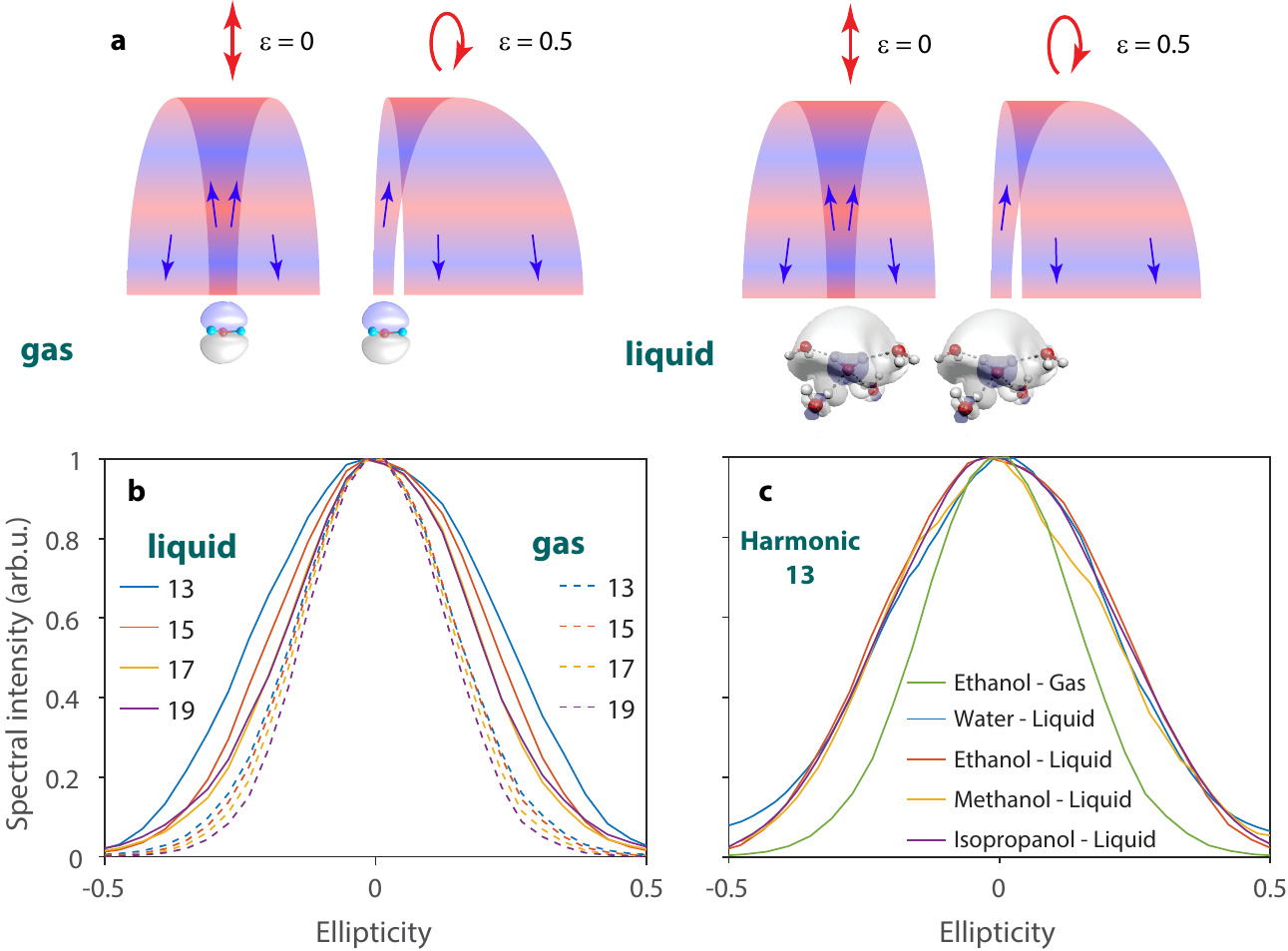}
\end{center}
\caption{{\bf Ellipticity dependence of liquid- vs- gas-phase HHG}. a) Schematic representation of the effect of ellipticity on the propagation of the electron wave packet. b) Dependence of the yield of selected harmonics on the ellipticity of the driving field in liquid and gaseous ethanol. c) Comparison of the ellipticity dependence of H13 from different liquids to gaseous ethanol. All measurements were recorded with a $\sim$30~fs driver, centered at 1.5~$\mu$m. Adapted from Ref. \cite{luu18a}.} 
\label{LHHG-ellip}
\end{figure}

Figure \ref{LHHG-ellip} compares the ellipticity dependence of the HHG yield from liquid- and gas-phase ethanol. The ellipticity dependence in general contains important information about the properties of the medium and the strong-field-driven electron dynamics. We find that the ellipticity dependence of all harmonic orders emitted from liquid ethanol is clearly broadened compared to the gas-phase emission. In the gas phase, the ellipticity dependence narrows down with increasing harmonic order, which is a signature of the laser-driven electron dynamics in the continuum: the longer the trajectory of the electron is, the more its trajectory is influenced by the laser field, which results in a larger sensitivity to ellipticity. The same observation applies to the liquid-phase ellipticity dependence, which also narrows down with increasing harmonic order. This observation suggests that a trajectory-based understanding of high-harmonic generation in liquids might be adequate. Figure \ref{LHHG-ellip}c compares the ellipticity dependence of H13 from water and several alcohols to that of gas-phase ethanol. The ellipticity dependencies of all liquids are very similar and considerably broader than those of the gas phase. The broadening of the ellipticity dependence in the liquid phase can have several origins, which include i) differences in the continuum-electron propagation, ii) electron scattering in liquids, iii) a different spatial extension of the electron-hole wavefunctions. Explanation i) is insufficient to rationalize the large observed effects because the differences in the ionization energies between the isolated and condensed molecules are on the order of 1~eV and the electron mass is not noticeably reduced for electron propagation in the liquids. Explanation ii) is more difficult to quantify because electron elastic mean-free paths are not available for alcohols and only recently became available in liquid water (see section \ref{MFP} and Ref. \cite{schild20a}). In Ref. \cite{luu18a}, we have used the mean-free paths determined for amorphous ice \cite{michaud03a}, for a lack or reliable data on liquid water. On this basis, we have assumed that electron scattering was negligible because the total propagation length of the electron trajectory emitting the cut-off harmonic (20~eV photon energy, emitted by a 1.5~$\mu$m driver with a peak electric field of 1.5~V/\AA) amounts to $\sim$3~nm, which is shorter than the mean-free paths including all types of collisions ($\sim$4~nm) at the corresponding kinetic energy. However, using the elastic mean-free path determined in our recent work (see Section \ref{MFP}), which amounts to $\sim$0.6~nm, we now conclude that elastic scattering could contribute, at least to some extent, to the observed broadening of the ellipticity dependence. An important reason for our EMFP being much longer is that it is based on a physical DCS, which was lacking in the analysis of the amorphous-ice data \cite{michaud03a}. The possible contributions of elastic scattering to the ellipticity dependence will be studied in the future using a trajectory-based model of liquid-phase HHG that we sketch in Section \ref{LHHG-theory}. Finally, effect iii) definitely contributes to, and probably dominates, the observed broadening of the ellipticity dependence. In the gas phase, the continuum electron wave packet has to return to the parent molecule for high-harmonic emission to take place. This causes the width of the ellipticity dependence to reflect the spatial extension of the ionized orbital(s). It is known from many calculations (see e.g. Refs. \cite{prendergast05a}) that the outermost valence orbitals of liquid water (and alcohols) are delocalized over several molecules. The delocalization over the first solvation shell is strong, but further delocalization is rather weak as a consequence of disorder in liquids. This delocalization is illustrated in Fig. \ref{LHHG-ellip}a, which shows the highest-occupied molecular orbital of a water pentamer, characteristic of a water molecule with one complete solvation shell. This significant delocalization of the electron hole leads to a broadening of the ellipticity dependence, which will contribute significantly to the observed effects. A quantitative verification of the relative importance of the discussed effects will become possible in the near future by adapting the methodological developments described in Section \ref{concepts} to liquid-phase HHG, as briefly discussed below (Section \ref{LHHG-theory}).

\subsection{Concepts and theoretical methods}
\label{LHHG-theory}

The theoretical description of HHG in liquids meets with several challenges. First, liquids have a density comparable to solids, such that the typical extensions of continuum-electron trajectories known from gas-phase HHG would correspond to the electron wave packet encountering several neighboring molecules on its trajectory. Consequently, the methods developed to understand HHG in solids might be the better starting point. However, and this is the second challenge, liquids are intrinsically disordered, which prevents the rigorous application of a momentum-space description of HHG, which has enabled rapid progress in the understanding of HHG in solids. Nevertheless, the absence of long-range order does not completely exclude the application of momentum-space methods because liquids possess short-range order. As long as the spatial extension of the continuum-electron trajectory is comparable to the length scale of this short-range order, the latter can be expected to play a role in HHG.

Based on these considerations, we have chosen the semiconductor Bloch equations (SBE) as a starting point for modelling liquid-phase HHG. The electronic structures of liquids are indeed commonly understood as those of large-bandgap semiconductors \cite{cabral05a}. The presence of local order, which is enhanced by the existence of strong hydrogen bonds, additionally justifies the use of an effective band structure \cite{prendergast05a}. The advantage of the SBE is that they naturally include both interband and intraband contributions to HHG, such that additional assumptions regarding the mechanisms at play are not needed \cite{luu16a}. The solution of the SBE with a realistic band structure, such as that derived from a calculation of a slab of e.g. 128 molecules however quickly reaches the limit of computational feasibility, in addition to the limits imposed by the accuracy of density-functional-theory calculations.

We have therefore chosen, as a zero-order approximation, an approach that circumvents all of these challenges. We have used the density of occupied and unoccupied states, measured by state-of-the-art X-ray spectroscopies, to derive a highly simplified model band structure that reflects the known densities of states, arising from the three occupied outer-most valence bands and the two lowest-lying unoccupied conduction bands. The main goals of this approach were to identify i) the sensitivity of the calculated HHG spectra to the electronic structure of the liquids, i.e. the band gap and the width of the bands, ii) the number of contributing bands, and iii) the capability of such a simple model to explain the characteristic differences in the HHG spectra of liquid water as compared to the alcohols.

\begin{figure}[h!]
\begin{center}
\includegraphics[width=\textwidth]{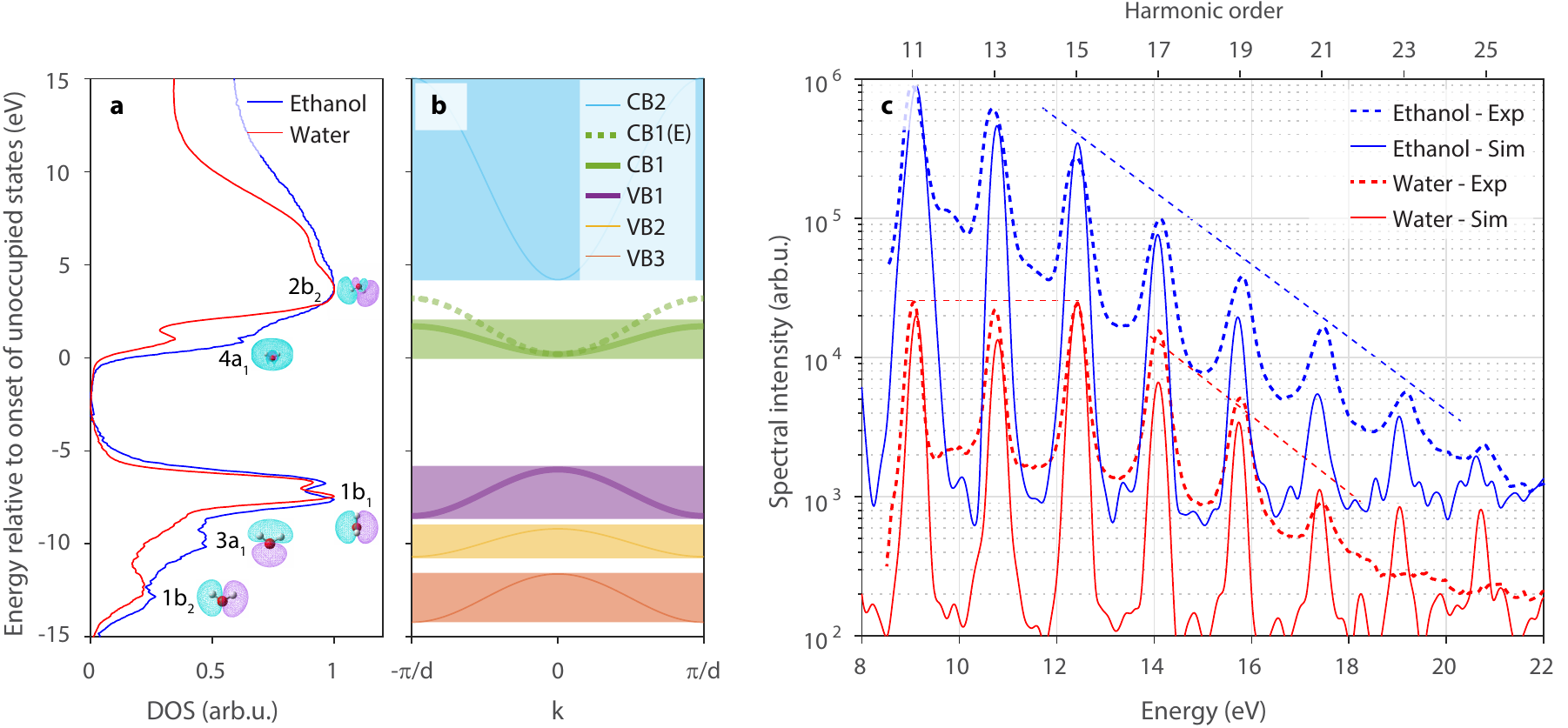}
\end{center}
\caption{{\bf Sensitivity of HHG to the electronic structure of liquids}. a) Densities of states of liquid water and ethanol obtained form X-ray spectroscopy, b) model ''band structure'' chosen to reflect these densities of states, c) measured HHG spectra of liquid water and ethanol using a $\sim$30 fs laser pulse centered at 1.5~$\mu$m and spectra calculated by solving the SBE for the model ''band structure'' shown in b. Adapted from Ref. \cite{luu18a}.} 
\label{LHHG-model}
\end{figure}

Figure \ref{LHHG-model}a shows the densities of states (DOS) of liquid water and ethanol as measured by X-ray spectroscopy \cite{tokushima08a,fuchs08a,schreck14a,schreck16a}. In the case of water, these consist of three outer-valence bands, which are labelled according to the symmetries of the orbitals of the isolated water molecule (1b$_1$, 3a$_1$ and 1b$_2$). The unoccupied states consist of two conduction bands, that are labeled according to the symmetries of the unoccupied orbitals of the isolated water molecule (4a$_1$ and 2b$_2$). The DOS of ethanol is comparable to that of water, with an important difference being the absence of a local maximum at the position of the 4a$_1$ band of liquid water. Based on these DOS and the results of detailed ''band-structure'' calculations \cite{prendergast05a}, we have derived the highly simplified model band structure shown in Fig. \ref{LHHG-model}b. The difference in the DOS of liquid water and ethanol is accounted for by choosing a larger width for the lowest conduction band of liquid ethanol compared to liquid water.

The solution of the SBE using these model band structures yields high-harmonic spectra that are in very good agreement with the observed spectra (full and dashed lines, respectively, in Fig. \ref{LHHG-model}). Most importantly, these calculations fully reproduce the main characteristics of the high-harmonic spectra, i.e. the monotonic decrease in the spectrum of ethanol and the plateau observed at the lowest three harmonic orders in liquid water, followed by a cut-off. We have verified that the calculated spectra do not change significantly when the two lower-lying valence bands and the highest-lying conduction band are removed from the calculation. Therefore the calculated high-harmonic spectra are mainly sensitive to the properties of the highest-lying valence and the lowest-lying conduction band.

Within our model, the characteristic differences between the HHG spectra of water and ethanol therefore exclusively originate from the different widths of the conduction bands, for which a cosinusoidal shape was assumed. The larger width in the case of ethanol causes the monotonic decay of the HHG spectrum, whereas the narrower width in the case of water causes the appearance of the plateau and cutoff. An additional characteristic of the HHG spectrum of liquid water is the appearance of a local minimum at H13 (10.7 eV). This local minimum is also reproduced in the SBE calculations. As we showed in Fig. 7 of Ref. \cite{luu18a}, the position of this minimum shifts in energy according to the size of the bandgap.

The comparison between our measurements and calculations therefore shows that high-harmonic spectroscopy of liquids is sensitive to the electronic structure of liquids, particularly to the properties of the conduction band, as well as the bandgap.

In the future, the methods for describing liquid-phase HHG can be improved in different ways. The most accurate computationally tractable approach would probably consist in running real-time, real-space time-dependent density-functional-theory (TDDFT) calculations on slabs of liquids sampled from molecular-dynamics calculations. This approach has the advantage of requiring minimal assumptions beyond those inherent to the TDDFT approach. It has the disadvantage that the physical insight into the mechanisms at play remains limited. Therefore, it might be desirable to develop complementary approaches, which offer additional insights. One such approach would consist in Monte-Carlo trajectory calculations of the strong-field driven electron dynamics. These calculations could rely on the principles described in Section \ref{concepts} and thereby incorporate electron scattering and laser-assisted electron scattering. The additionally required strong-field-ionization and photorecombination matrix elements could be obtained by extending the methods established for isolated molecules to the condensed phase by using the cluster approach described in Section \ref{concepts}. This method would offer additional insights into liquid-phase HHG because the role of electron scattering, orbital delocalization and many other effects could be disentangled. The challenge for this approach will be to achieve a sufficient accuracy to reach agreement with the experimental data.

The results presented in this section set the foundations for the development of liquid-phase high-harmonic spectroscopy. In the future, the detailed mechanism of liquid-phase HHG will be studied with attosecond temporal resolution by using both {\it in-situ} methods such as two-color HHG \cite{shafir12a,vampa15a} and attosecond transient absorption \cite{smith20a} based on water-window high-harmonic sources \cite{pertot17a} or {\it ex-situ} methods, such as attosecond photoelectron spectroscopy \cite{paul01a}. Of particular interest are the capabilities of HHS to resolve attosecond electron dynamics in the liquid phase, such as electron-liquid scattering dynamics and the spatial characteristics of electron-hole dynamics in liquids. Targets of primary interest will be water itself, but also aqueous solutions, which offer the possibility to study solvated species in their natural environment, and other liquids.

\section{Conclusions and Outlook}

In this chapter, we have described two novel experimental approaches that enable attosecond time-resolved experiments to be performed on liquid samples. The crucial innovations have been the development of attosecond photoelectron spectroscopy with cylindrical liquid microjets and the demonstration of high-harmonic spectroscopy with flat microjets. 

In the first case, a general novel approach to analyze and interpret overlapping attosecond photoelectron spectra has been developed. This technique generalizes attosecond photoelectron spectroscopy not only to the liquid phase, but actually to any complex sample. On the conceptual side, a general theoretical framework for condensed-phase attosecond photoelectron spectroscopy has been developed that includes, for the first time, the treatment of photoionization and scattering delays, as well as a coherent treatment of all processes involved. This framework has been validated first by benchmarking against the time-dependent Schr\"odinger equation and finally by reaching quantitative agreement with attosecond time delays of liquid water.

In the second case, a general approach to high-harmonic spectroscopy in liquids has been established. This result represents a change of paradigms in a field where XUV high-harmonic generation from liquids was previously assumed to be impossible. Our novel experimental approach has enabled the first observation of XUV HHG from liquids, the unequivocal separation and simultaneous measurement of liquid- and gas-phase HHG and their detailed comparison. We have found a linear scaling of the high-harmonic cut-off energy with the peak electric field of the driver and a highly non-perturbative scaling of the HHG yield. We have found a systematically and substantially broadened dependence of the HHG yields on the ellipticity of the driving field, compared to the gas phase. Finally, based on the solution of the semiconductor Bloch equations of a strongly-driven model band system, we have found a pronounced sensitivity of the HHG spectra on the electronic structure of liquids, particularly the properties of the conduction band and the band gap.

These developments establish two major avenues for developing attosecond time-resolved spectroscopy of liquids and solutions. They open a myriad of future possibilities to explore the role of electronic dynamics in solvated atoms, ions, molecules and nanoparticles in their natural environment. Of particular interest for such work are the prototypical processes of charge and energy transfer. We will briefly discuss two examples of such processes. Intermolecular Coulombic decay \cite{cederbaum97a} in liquid water is the prototype for energy transfer in the liquid phase and a source of slow electrons in liquid water. ICD occurs when the inner-valence (2a$_1$) or the core (1a$_1$) orbitals of water (clusters or liquid) are ionized. So far, ICD has only been observed in water dimer \cite{jahnke10a} and water clusters \cite{mucke10a,richter18a}, but not in bulk liquid water. Time-resolved experiments on ICD in water are equally missing. Recent theoretical work predicts a time scale of 12-52 fs for ICD in small water clusters \cite{richter18a}. However, a different type of calculations predicts ICD lifetimes of 3.6-4.6~fs for isoelectronic (HF)$_3$ clusters \cite{ghosh14a}, suggesting that the ICD lifetimes in water (clusters) could also be much shorter.
Electron-transfer-mediated decay \cite{zobeley01a} is a prototype of ultrafast charge transfer reactions in liquids. ETMD occurs when the inner-valence or core hole created by ionization is filled by an electron from a neighboring particle. The energy made available in this process then serves to eject a second electron from the particle that provided the first electron to fill the hole (ETMD(2)) or a third particle (ETMD(3)). ETMD has only recently been observed in the liquid phase \cite{unger17a}. Its time scale is presently unknown. 

These two examples illustrate the possible future of liquid-phase attosecond science. They represent two relatively simple and very fundamental mechanisms of electronic dynamics in liquids. They are representative of a broad variety of electronic processes that play a role in most chemical reactions and biological transformations. Looking forward, the methods described in this chapter will provide access to the study of many important, but poorly understood electronic processes in the liquid phase.

\section*{Acknowledgements}
We gratefully acknowledge the contributions of many co-workers and collaborators who have contributed to this work over several years.
This work was financially supported by an ERC starting grant (project no. 307270-ATTOSCOPE), an ERC consolidator grant (project no. ATTOLIQ) and the Swiss National Science Foundation (SNSF) via the National Center of Competence in Research Molecular Ultrafast Science and Technology and projects no. 200021\_159875 and 200021\_172946. 
A. S. is grateful for financial support from an Ambizione grant of the  SNSF. D. J. thanks the FP-RESOMUS fellowship program.
%\end{acknowledgments} 

%\bibliographystyle{spphys}
%\bibliography{../../attobib}

\end{document}